\documentclass[sigconf,9pt]{acmart}
\usepackage[english]{babel}

\usepackage{blindtext}

\usepackage{tikz}
\usepackage{amsmath}

\usepackage{filecontents}

\usepackage{tikz}
\usepackage{xcolor}
\usepackage{xspace}
\usepackage{tikz}
\usepackage{pifont}
\usepackage[english]{babel}
\usepackage{blindtext}
\usepackage{lipsum}
\usepackage{colortbl} 
\usepackage[ruled,vlined,linesnumbered]{algorithm2e}
\usepackage{booktabs}

\usepackage{filecontents}
\usepackage{xspace}
\usepackage{subfig}
\usepackage{xcolor}
\usepackage{hhline}
\usepackage{multirow}
\usepackage{soul}

\usepackage{array}
\usepackage{comment}
\usepackage{listings}
\usepackage{dsfont}
\usepackage{algorithmic}
\usepackage{xurl}
\usepackage{setspace}
\usepackage{graphicx}
\usepackage{caption}
\usepackage{ltablex}

\usepackage{ifluatex}

\usepackage{outlines}

\newtheorem{insight}{{\bf Insight}}
\setcopyright{acmlicensed}

\acmYear{2024}
\copyrightyear{2024}
\acmConference[ACM SIGCOMM '24]{ACM SIGCOMM 2024 Conference}{August 4--8, 2024}{Sydney, NSW, Australia}
\acmBooktitle{ACM SIGCOMM 2024 Conference (ACM SIGCOMM '24), August 4--8, 2024, Sydney, NSW, Australia}
\acmDOI{10.1145/3651890.3672274}
\acmISBN{979-8-4007-0614-1/24/08}

\begin{document}
\pagenumbering{gobble}



\newcommand*{\affmark}[1][*]{\textsuperscript{#1}}
\newcommand*{\affaddr}[1]{#1}

\title{\name: KV Cache Compression and Streaming for Fast Large Language Model Serving 
}

\author{\Large { Yuhan Liu, Hanchen Li, Yihua Cheng, Siddhant Ray, Yuyang Huang, Qizheng Zhang*, Kuntai Du, Jiayi Yao, \\Shan Lu\affmark[$\dag$], Ganesh Ananthanarayanan\affmark[$\dag$], Michael Maire, Henry Hoffmann, Ari Holtzman,  Junchen Jiang} \\
\affaddr{\textit{University of Chicago}}~~~~\affaddr{\affmark[$\dag$]{\textit{Microsoft}}~~~~\affaddr{*{\textit{Stanford University}}} 
}
}
\renewcommand{\shortauthors}{Y. Liu, et al}


\soulregister\sf7


\newcommand{\yhedit}[1]{{\color{black} #1}}

\newcommand{\edit}[1]{{\color{black} #1}}
\newcommand{\shan}[1]{{\color{brown}(Shan: #1)}}
\newcommand{\shanedit}[1]{{\color{red} #1}} 
\newcommand{\hank}[1]{{\color{blue}(Hank: #1)}}
\newcommand{\mm}[1]{{\color{violet}(Michael:#1)}}
\newcommand{\jc}[1]{{\footnotesize\color{blue}{(JC: #1)}}}
\newcommand{\jcedit}[1]{{\color{orange} #1}} 
\newcommand{\yh}[1]{{\footnotesize\color{deepgreen}{(Yuhan: #1)}}}
\newcommand{\todo}[1]{{\color{red}{(TODO: #1)}}}
\definecolor{darkkhaki}{rgb}{0.74, 0.72, 0.42}
\newcommand{\hc}[1]{{\color{darkkhaki}{(LHC: #1)}}}
\newcommand{\ga}[1]{{\color{cyan!70!blue}{(Ganesh: #1)}}}
\newcommand{\sid}[1]{{\footnotesize \color{brown!60!blue}{(Sid: #1)}}}
\definecolor{yccolor}{RGB}{240, 24, 240}
\newcommand{\yc}[1]{{\color{yccolor}{\footnotesize{[Yihua: #1]}}\xspace}}






\newcommand{\NumTokens}{\ensuremath{{N}}\xspace}
\newcommand{\NumChannels}{\ensuremath{{c}}\xspace}
\newcommand{\NumLayers}{\ensuremath{{l}}\xspace}
\newcommand{\Level}{\ensuremath{{k}}\xspace}

\newcommand{\Decision}{Decision\xspace}
\newcommand{\x}{\ensuremath{\mathbf{x}}\xspace}
\newcommand{\y}{\ensuremath{\mathbf{y}}\xspace}
\newcommand{\z}{\ensuremath{z}\xspace}
\newcommand{\X}{\ensuremath{\mathbf{X}}\xspace}
\newcommand{\TestNum}{\ensuremath{T}\xspace}
\newcommand{\APIcomp}{App \circ Filter\xspace}
\newcommand{\Filter}{F\xspace}
\newcommand{\DNN}{DNN\xspace}
\newcommand{\API}{API\xspace}
\newcommand{\Appdecision}{App\xspace}
\newcommand{\total}{\ensuremath{t}\xspace}
\newcommand{\NTarget}{\ensuremath{N}\xspace}
\newcommand{\other}{\ensuremath{O}\xspace}

\newcommand{\Rate}{IDR\xspace}

\newcommand{\T}{\ensuremath{T}\xspace}
\newcommand{\Match}{\ensuremath{M}\xspace}
\newcommand{\BeforeBreak}{\ensuremath{B}\xspace}

\newcommand{\App}{$A$\xspace}
\newcommand{\M}{\ensuremath{M}\xspace}
\newcommand{\F}{\ensuremath{\alpha}\xspace}
\newcommand{\G}{\ensuremath{\beta}\xspace}
\newcommand{\Q}{\ensuremath{\gamma}\xspace}
\newcommand{\ClassId}{\ensuremath{c}\xspace}
\newcommand{\LabelId}{\ensuremath{l}\xspace}
\newcommand{\InputId}{\ensuremath{i}\xspace}
\newcommand{\TraverseId}{\ensuremath{j}\xspace}
\newcommand{\score}{effective score\xspace}
\newcommand{\scores}{effective scores\xspace}
\newcommand{\Tg}{\ensuremath{TC}\xspace}
\newcommand{\TargetSet}{\ensuremath{G}\xspace}
\newcommand{\MaxScore}{\ensuremath{f(\y)}\xspace}

\newcommand{\summary}{decision-process summary\xspace}
\newcommand{\Summary}{Decision-process summary\xspace}
\newcommand{\summaries}{decision-process summaries\xspace}
\newcommand{\target}{target class\xspace}
\newcommand{\targets}{target classes\xspace}
\newcommand{\Target}{Target class\xspace}
\newcommand{\process}{software decision process\xspace}
\newcommand{\Process}{Software decision process\xspace}
\newcommand{\bestapi}{Best-of-all API*\xspace}
\newcommand{\qa}{interactive Q \& A\xspace}
\newcommand{\rag}{retrieval-augmented generation\xspace}
\newcommand{\kv}{key-value cache\xspace}
\newcommand{\scis}{Scissorhands\xspace}
\newcommand{\recomp}{Loading text context\xspace}

\newcommand{\name}{CacheGen\xspace}
\newcommand{\modeld}{$\textrm{ChameleonAPI}_{basic}$\xspace}
\newcommand{\tool}{ChameleonAPI\xspace}
\newcommand{\toolbasic}{\ensuremath{\textrm{ChameleonAPI}_{basic}}\xspace}
\newcommand{\code}[1]{{\texttt{#1}}}
\newcommand{\term}[1]{\textsf{#1}}
\newcommand{\wl}{\emph{whitelist}\xspace}
\newcommand{\wls}{\emph{whitelists}\xspace}
\newcommand{\ift}{\emph{if/then branch}\xspace}
\newcommand{\ifts}{\emph{if/then branches}\xspace}

\newcommand{\fillme}{{\bf XXX}\xspace}

\newcommand*\circled[1]{\tikz[baseline=(char.base)]{
            \node[shape=circle,fill,inner sep=2pt] (char) {\textcolor{white}{\footnotesize{#1}}};}}

\newcounter{packednmbr}
\newenvironment{packedenumerate}{\begin{list}{\thepackednmbr.}{\usecounter{packednmbr}\setlength{\itemsep}{0.5pt}\addtolength{\labelwidth}{-4pt}\setlength{\leftmargin}{2ex}\setlength{\listparindent}{\parindent}\setlength{\parsep}{1pt}\setlength{\topsep}{0pt}}}{\end{list}}
\newenvironment{packeditemize}{\begin{list}{$\bullet$}{\setlength{\itemsep}{0.5pt}\addtolength{\labelwidth}{-4pt}\setlength{\leftmargin}{2ex}\setlength{\listparindent}{\parindent}\setlength{\parsep}{1pt}\setlength{\topsep}{2pt}}}{\end{list}}
\newenvironment{packedpackeditemize}{\begin{list}{$\bullet$}{\setlength{\itemsep}{0.5pt}\addtolength{\labelwidth}{-4pt}\setlength{\leftmargin}{\labelwidth}\setlength{\listparindent}{\parindent}\setlength{\parsep}{1pt}\setlength{\topsep}{0pt}}}{\end{list}}
\newenvironment{packedtrivlist}{\begin{list}{\setlength{\itemsep}{0.2pt}\addtolength{\labelwidth}{-4pt}\setlength{\leftmargin}{\labelwidth}\setlength{\listparindent}{\parindent}\setlength{\parsep}{1pt}\setlength{\topsep}{0pt}}}{\end{list}}
\let\enumerate\packedenumerate
\let\endenumerate\endpackedenumerate
\let\itemize\packeditemize
\let\enditemize\endpackeditemize

\newcommand{\tightcaption}[1]{\vspace{-0.15cm}\caption{{\normalfont{\textit{{#1}}}}}\vspace{-0.3cm}}
\newcommand{\tightsection}[1]{\vspace{-0.1cm}\section{#1}\vspace{-0.00cm}}
\newcommand{\tightsectionstar}[1]{\vspace{-0.17cm}\section*{#1}\vspace{-0.08cm}}
\newcommand{\tightsubsection}[1]{\vspace{-0.1cm}\subsection{#1}\vspace{-0.02cm}}
\newcommand{\extightsubsection}[1]{\vspace{-0.3cm}\subsection{#1}\vspace{-0.02cm}}
\newcommand{\tightsubsubsection}[1]{\vspace{-0.01in}\subsubsection{#1}\vspace{-0.01cm}}

\newcommand{\eg}{{\it e.g.,}\xspace}
\newcommand{\ie}{{\it i.e.,}\xspace}
\newcommand{\etal}{{\it et.~al}\xspace}
\newcommand{\bigO}{\mathrm{O}}
\newcommand{\twlog}{w.l.o.g.\xspac}

\newcommand{\myparashort}[1]{\vspace{0.05cm}\noindent{\bf {#1}}~}
\newcommand{\mypara}[1]{\vspace{0.05cm}\noindent{\bf {#1}:}~}
\newcommand{\myparatight}[1]{\vspace{0.02cm}\noindent{\bf {#1}:}~}
\newcommand{\myparaq}[1]{\vspace{0.05cm}\noindent{\bf {#1}?}~}
\newcommand{\myparaittight}[1]{\smallskip\noindent{\emph {#1}:}~}
\newcommand{\question}[1]{\smallskip\noindent{\emph{Q:~#1}}\smallskip}
\newcommand{\myparaqtight}[1]{\smallskip\noindent{\bf {#1}}~}

\newcommand{\cmark}{\ding{51}}%
\newcommand{\xmark}{\ding{55}}%



\definecolor{backcolour}{rgb}{0.96,0.96,0.96}
\definecolor{codegray}{rgb}{0.5,0.5,0.5}
\definecolor{deepblue}{rgb}{0,0,0.6}
\definecolor{deepred}{rgb}{0.6,0,0}
\definecolor{deepgreen}{rgb}{0,0.5,0}
\lstdefinestyle{mystyle}{
    backgroundcolor=\color{backcolour},   
    commentstyle=\color{codegreen},
    morekeywords={self, True},
    keywordstyle=\color{deepblue},
    numberstyle=\tiny\color{codegray},
    emph={MyClass,__init__,EncodingType,Image},
    emphstyle=\color{deepred},
    stringstyle=\color{deepgreen},
    basicstyle=\ttfamily\footnotesize,
    breakatwhitespace=false,         
    breaklines=true,                 
    captionpos=b,                    
    keepspaces=true,                 
    numbers=left,                    
    numbersep=5pt,                  
    showspaces=false,                
    showstringspaces=false,
    showtabs=false,                  
    tabsize=1
}

\newcommand{\mycaption}[3]{
\begin{spacing}{0.95}
\caption{
\label{#1}
{\bf #2. }
{\it \small #3}
}
\end{spacing}
}

\newcommand{\vminfive}{\vspace{-5pt}}
\newcommand{\vminten}{\vspace{-10pt}}
\newcommand{\vminfifteen}{\vspace{-15pt}}
\newcommand{\vmintwenty}{\vspace{-20pt}}
\newcommand{\vmintwentyfive}{\vspace{-25pt}}

\def \hmina {\hspace{-0.1in}}
\def \hminb {\hspace{-0.2in}}


\begin{abstract}

As large language models (LLMs) take on complex tasks, their inputs are supplemented with {\em longer contexts} that incorporate domain knowledge.
Yet using long contexts is challenging
as nothing can be generated until the whole context is processed by the LLM. While the context-processing delay can be reduced by reusing the KV cache of a context across different inputs, fetching the KV cache, which contains large tensors, over the network can cause high extra network delays.

\name is a fast context-loading module for LLM systems. First, \name uses a custom tensor encoder, leveraging KV cache's distributional properties to {\em encode}  a KV cache into more compact bitstream representations with negligible decoding
overhead, to save bandwidth usage.
Second, \name{ }{\em adapts} the compression level of different parts of a KV cache to cope with changes in available bandwidth, in order to maintain low context-loading delay and high generation quality.
We test \name on popular LLMs and datasets. Compared to the recent systems
that reuse the KV cache, CacheGen reduces the KV cache size by
3.5-4.3x and the total delay in fetching and processing contexts
by 3.2-3.7x with negligible impact on the LLM response
quality.
Our code is at: \href{https://github.com/UChi-JCL/CacheGen}{\textbf{https://github.com/UChi-JCL/CacheGen}}.

\end{abstract}

\maketitle

\vspace{-6pt}
{\par\smallskip\small\noindent{\bfseries ACM Reference Format:}\par\nobreak
  \noindent Yuhan Liu, Hanchen Li, Yihua Cheng, Siddhant Ray, Yuyang Huang, Qizheng Zhang, Kuntai Du, Jiayi Yao, Shan Lu, Ganesh Ananthanarayanan, Michael Maire, Henry Hoffmann, Ari Holtzman,  Junchen Jiang. 2024. \name: KV Cache Compression and Streaming for Fast Large Language Model Serving.
  In \textit{SIGCOMM '24, August 4–August 8, 2024, Sydney, Australia. ACM, New York, NY, USA, 18 pages}
  }



\tightsection{Introduction}

With impressive generative quality,
large language models (LLMs) are ubiquitously used~\cite{llm-app-1,llm-app-2,llm-app-3,llm-app-4} in personal assistance, AI healthcare, and marketing. 
The wide use of LLM APIs (\eg OpenAI GPT-4~\cite{gpt4-api}) and the industry-quality open-source models (\eg Llama~\cite{llama}), combined with popular application frameworks (\eg HuggingFace~\cite{huggingface_rag}, Langchain~\cite{langchain}), further boosts LLMs' popularity. 

To perform complex tasks, users or applications often prepend an LLM input 
with a {\em long context} containing thousands of tokens or more. 
For example, some context supplements user prompts with
domain-knowledge text so that the LLM can generate responses using specific knowledge
not embedded in the LLM itself.
As another example, a user prompt can be supplemented with the conversation histories
accumulated during the interactions between the user and the LLM.
Though short inputs are useful~\cite{arxiv-1,arxiv-2}, 
longer inputs often improve response quality and coherence~\cite{beltagy2020longformer,long-llama, wu2022memorizing, dai*2019transformerxl,roy2021efficient,bertsch2023unlimiformer, fid, retro}, which has fueled the ongoing race to train LLMs that accept ever longer inputs, 
from 2K tokens in ChatGPT to 100K in Claude~\cite{long-context}. 


Using long contexts poses a challenge to the response generation {\em latency}, as no response can be generated until the whole context is loaded and processed by the LLM.
The amount of computation in processing a long context grows super-linearly with the context length~\cite{vaswani2023attention, dao2022flashattention, beltagy2020longformer, zaheer2021big, roy2021efficient}.
While some recent works increase the throughput of processing long context~\cite{agrawal2023sarathi}, the {\em delay} of processing the context can still be several seconds for long contexts (2 seconds for a 3K context)~\cite{agrawal2023sarathi, promptcache}.
In response, many systems reduce the context-processing delay by storing and reusing the {\em KV cache} of the context to skip redundant computation when the context is used again (\eg~\cite{vllm, chunkattention, promptcache, sglang}).

\begin{figure}[t]
\centering
     \includegraphics[width=.99\linewidth]{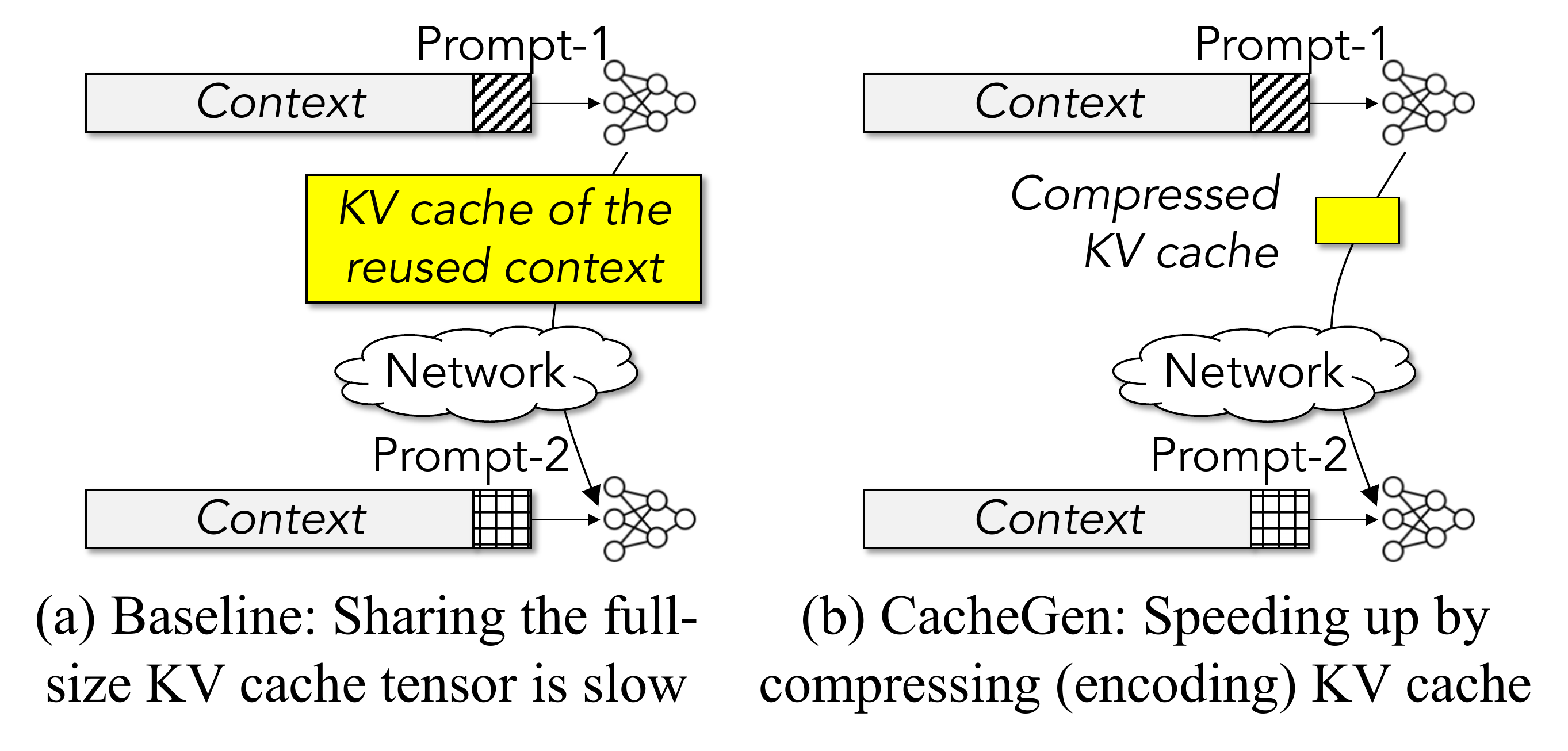}
     \vspace{-10pt}
\tightcaption{When the  context is reused, \name speeds up the sharing of its KV cache by compressing (encoding) the KV cache.} 
    \label{fig:intro}
    \vspace{-8pt}
\end{figure}

Yet, the KV cache of a reused context may {\em not} always be in local GPU memory when the next input comes; instead, the KV cache may need to be retrieved from another machine(s) first, causing extra network delays (Figure~\ref{fig:intro}a).
For instance, a database of background documents might reside in a separate storage service, and the documents (\ie context) assisting LLM inference are only to be selected and fetched to the LLM when a relevant query is received~\cite{langchain-retrival,arxiv-information-retrieval,generative-agents,autogpt,beltagy2020longformer}.

The extra network delay for fetching the KV cache has not yet received much attention. 
Previous systems assume the KV cache of a context is always kept in the same GPU memory between different requests sharing the same context~\cite{promptcache}, or the KV cache is small enough to be sent quickly by a fast interconnection~\cite{zhong2024distserve, patel2023splitwise}.
Yet, as elaborated in \S\ref{sec:delay}, the delay for fetching a KV cache can be non-trivial,
since a KV cache consists of large high-dimensional floating-point tensors, whose sizes grow with both the context length and model size and can easily reach 10s~GB.
The resulting network delay can be 100s milliseconds to over 10 seconds, hurting the interactive user experience~\cite{latency_experience1, latency_experience2, latency_experience3}.
In short, when loading contexts' KV cache from other machines, solely optimizing computational delay may cause {\em higher} response latency, as loading the KV cache increases the network delay.

There have been a few recent efforts to reduce the \emph{run-time} size of KV cache in GPU memory in order to fit the memory limit or LLM's input limit.
Some drop unimportant tokens from KV cache or context text~\cite{scissorhand, zhang2023h2o, jiang2023llmlingua, jiang2023longllmlingua}, and others apply smart quantization on KV cache tensor~\cite{kang2024gear, liu2024kivi, hooper2024kvquant}. 
In contrast, we want to reduce the \emph{transmission-time} size of KV cache to reduce the \emph{network delay}.
Thus, we do {\em not} need to keep the tensor format of KV cache and, instead, can encode it into more compact bitstreams.

We present \name, a fast context-loading module in LLM systems for reducing the network delay in fetching and processing long contexts (Figure~\ref{fig:intro}b).
It entails two techniques.




\mypara{KV cache {\em encoding and decoding}}
\name encodes a precomputed KV cache into more compact {\em bitstream} representations, rather than keeping the tensor shapes of the KV cache.
This greatly saves bandwidth and delays when sending a KV cache.
Our KV cache encoder employs a custom quantization and arithmetic coding strategy to leverage the distributional properties of KV cache, such as locality of KV tensors across nearby tokens and different sensitivities towards quantization losses at different layers of a KV cache.
Furthermore, the decoding (decompression) of KV caches is accelerated by a GPU-based implementation, and the decoding is {\em pipelined} with transmission to further reduce its impact on the overall inference delay.

\mypara{KV cache {\em streaming}}
\name streams the encoded bitstreams of a KV cache in a way that adapts to changes in network conditions.
Before a user query arrives, \name splits a long context into chunks and encodes the KV of each chunk separately at various compression levels (similar to video streaming). 
When sending a context's KV cache, \name 
fetches the chunks one by one and adapts the per-chunk compression level to maintain high generation quality while keeping the network delay within a Service-Level Objective (SLO). 
When the bandwidth is too low, \name can also fall back to sending a chunk in text format and leave it to the LLM
to recompute the KV cache of the chunk.

In short, unlike prior systems that optimize the KV cache in GPU memory, \name focuses on the {\em network} delay for sending the KV cache.
We compare \name with a range of baselines, including KV quantization~\cite{flexgen}, loading contexts in text form, and state-of-the-art context compression~\cite{jiang2023longllmlingua,zhang2023h2o}, using three popular LLMs of various sizes (from 7B to 70B) and four datasets of long contexts (662 contexts with 1.4~K to 16~K tokens). Table~\ref{tab:snapshot} gives a preview of the results. 
Our key findings are:
\begin{itemize}
\item In terms of the delay of transmitting and processing contexts (\ie time-to-first-token), \name is 3.2-3.7$\times$ faster than the quantization baseline at the similar generation quality (F1 score and perplexity), and 3.1-4.7$\times$ faster than loading the text contexts with less than 2\% accuracy drop. 
 Notably, compared with 8-bit quantization, a nearly lossless KV cache compression, \name is still able to reduce the delay of loading context by 1.67-1.81$\times$.
\item In terms of the bandwidth usage for sending KV cache, \name achieves the same generation quality while using 3.5-4.3$\times$ less bandwidth than the quantization baseline.
\item When combined with the recent context compression methods~\cite{zhang2023h2o, jiang2023longllmlingua}, \name further reduces the bandwidth usage for sending their KV caches by 3.3-4.2$\times$. 

\end{itemize}
This work does not raise any ethical issues.

\begin{table}[t]
\centering
\begin{tabular}{lcc}
\hline
\multirow{2}{*}{\textbf{Technique}} & \textbf{KV cache size} & \textbf{Accuracy} \\
& \footnotesize{(in MB, lower the better)} & \footnotesize{(higher the better)} \\ \hline
8-bit quantization & 622 & 1.00 \\ 
\rowcolor{blue!30} \name (this paper) & 176  & 0.98 \\ 
H2O~\cite{zhang2023h2o} & 282 & 0.97 \\ 
\rowcolor{blue!30} \name on H2O & 71 & 0.97 \\
LLMLingua~\cite{jiang2023longllmlingua} & 492 & 0.94 \\ 
\rowcolor{blue!30}\name on LLMLingua  & 183 & 0.94 \\ \hline
\end{tabular}
\vspace{5pt}
\tightcaption{Performance of \name and the baselines on Mistral-7B with LongChat dataset~\cite{longchat2023}. Full results are shown in \S\ref{sec:eval}. }
\vspace{-10pt}
\label{tab:snapshot}
\end{table}

\tightsection{Background and Motivation}

\vspace{0.2cm}




\tightsubsection{Large language model basics}
\label{subsec:transformer}


Transformers~\cite{vaswani2023attention,brown2020language, palm} are the de facto models for most large language model (LLM) services.
At a high level, a transformer takes a sequence of input tokens\footnote{A ``token'' 
can be a punctuation, a word, or a part of a word. 
Tokenizing an input is much faster than the generation process.}
and generates a sequence of output tokens through two phases.

During the prefill phase, an attention neural network takes in the input token. Then each of the $l$ layers in the attention module produces two two-dimensional tensors, a key (K) tensor and a value (V) tensor. 
These K and V tensors contain information essential for LLM to utilize the context later. All the KV tensors across different layers are together called the {\em KV cache}.



During the generation phase, also called the decoding phase, the KV cache is used to compute the attention score between every
pair of tokens, which constitute the attention matrix, and generate output tokens in an autoregressive manner. 
For performance reasons, the KV cache, which has a large memory footprint \cite{vllm}, 
is usually kept in GPU memory during this phase and released afterward. 
Some emergent optimizations save and reuse the KV cache across different LLM requests, as we will explain shortly.

In all mainstream models, the compute overhead of the prefill phase grows superlinearly with the input length.
Since the prefill phase must be completed before generating the first output token, its duration is called {\em Time-to-First-Token} ({\em TTFT}).
This paper focuses on reducing TTFT during prefilling while not changing the decoding process.

\tightsubsection{Context in LLM input}
\label{subsec:context}

LLMs may generate
low-quality or hallucinated answers when the response requires knowledge not 
already embedded in the models. 
Thus, many LLM applications and users supplement the LLM input with additional texts, referred to as the {\bf context}~\cite{gao2024retrievalaugmented, lewis2021retrievalaugmented}.
The LLM can read the context first and use its in-context learning capability to generate high-quality responses.\footnote{An example of this process is retrieval-augmented generation (RAG), which uses a separate logic to select the context documents for a given query,
It is well-studied in natural-language literature and widely used in industry.} 

The contexts in LLM input can be used for various purposes. 

{\em (i)} a user question can be supplemented with a document about specific domain knowledge,
to produce better answers~\cite{chatgpt,rubin2023long, anyscalerag}, including 
using latest news to answer fact-checking inquiries \cite{perplexity, llmapp}, 
using case law or regulation documents to offer legal assistance
\cite{saleem_llm_lawyers, medium_lawyers}, etc.;
{\em (ii)} code analysis applications retrieve context from a code repository to answer questions or generate a summary about the repository~\cite{bairi2023codeplan,jain2023llmassisted, jimenez2023swebench},
and similarly financial companies use LLMs to generate summaries or answer questions based on detailed financial documents~\cite{bloomberg};
{\em (iii)} gaming applications use the description of a particular character as context so that the LLM can generate character dialogues or actions matching the character personality~\cite{generative-agents, wu2023deciphering, shi2023cooperation};
{\em (iv)} in few-shot learning, a set of question-answer pairs are used as context to teach the LLM to answer certain types of questions~\cite{10.1145/3551349.3559555,10.1145/3626111.3628183, 10.1145/3582688};
{\em (v)} in chatting apps, the conversational history with a user is often prepended as the context to subsequent user input to produce consistent and informed  responses~\cite{openai_forum_2023, quora_question_year}. 




We observe that in practice, contexts are often {\bf long} and often {\bf reused} to supplement different user inputs. 



\emph{Long} contexts are increasingly common in practice.
For example, those contexts discussed above, such as 
case law documents, financial documents, news articles, code files, and chat history accumulated
in a session,
easily contain
thousands of tokens or more. 
Intuitively, longer contexts are more likely to include the right information and hence
may improve the quality of the response.
Indeed, FiD~\cite{fid} shows that the accuracy increases from 40\% to 48\% when the context increases from 1K tokens to 10K.
Retro~\cite{retro} similarly shows that the generation quality (perplexity) improves significantly when the context increases from 6K tokens to 24K. 
\yhedit{This paper focuses on contexts such as conversation histories accumulated in a chat session, or a single document
input by the user to provide necessary information needed to accomplish the task. }


These long contexts are often {\em reused} by different inputs.
In the financial analysis example, 
consider two queries, ``write a short summary based on the company's earning report last quarter'' and ``what were the company's top sources of revenue in the last quarter'';
the same earning reports are likely to be supplemented to both queries as the contexts.
Similarly, the same law enforcement document or latest news article 
can be used to answer many different queries in legal assistant or fact-checking apps.
As another example, during a chat session, early chat content will keep getting reused as part of the context for every later chat input.


In short, longer contexts lead to higher prefill delays and hence longer TTFT, but since the same contexts are often reused, it is promising to reduce TTFT by caching the intermediate results (i.e., the KV cache) and hence avoid prefill recomputation. This solution has indeed been explored recently
\cite{vllm, chunkattention, promptcache} and shown its potential with just one caveat, which we discuss in the next section.

\vspace{-4pt}
\tightsection{The Hidden Network Bottleneck}
\label{sec:delay}

While reusing the KV cache of a long context could drastically reduce TTFT, this benefit comes with a catch---the reused KV cache must be in the local GPU memory in the first place~\cite{vllm, chunkattention, promptcache, sglang}.

\begin{figure}[t]
\centering
     \includegraphics[width=1.02\linewidth]{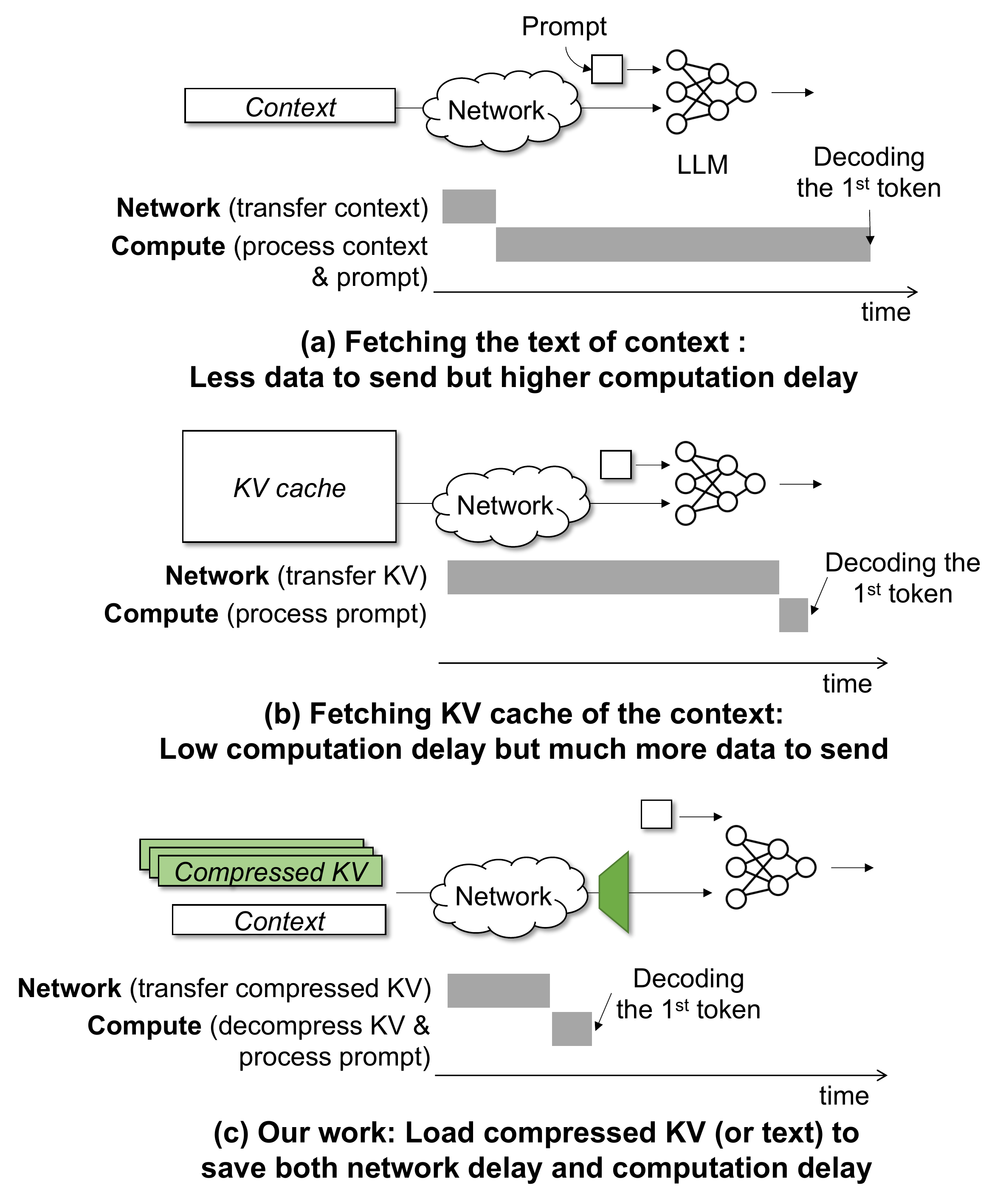}
     \vspace{-15pt}
    \tightcaption{How different ways of loading context affect the network delay (to transfer context or KV cache) and the computation delay (to run the attention module on the context).  } 
     \vspace{-6pt}
    \label{fig:motivate_baselines}
\end{figure}

\mypara{Why KV cache needs to be loaded}
In practice, however, the reused KV cache may need to be fetched from another machine(s).
This is because GPU memory is likely not enough to store the KV caches of many repeated contexts. 
\yhedit{
For example, in a financial assistance application, an LLM performs data analysis 
 on long financial reports~\cite{Nucci2024Large}, which can have thousands or tens of thousands of tokens, leading to a large KV cache size. 
To make it concrete, processing Amazon's annual report for 2023, which has $\sim$80,000 tokens~\cite{Amazon2023AnnualReport}, with the model of Llama-34B produces a KV cache of 19~GB, which is on par with the size of the LLM itself. 
As different queries that reuse a KV cache may be several hours apart, the reused KV cache may have to be offloaded to make space for fresh chat sessions.
Moreover, as newer LLMs can accept ever longer contexts~\cite{wu2024loongserve, lin2024infinitellm, ge2024incontext, ding2024longrope, hsieh2024ruler}, storing them on dedicated storage servers, rather than CPUs or GPU, would be more practical and economical.}
Besides, different requests that reuse KV cache may not always hit the same GPU, which also requires the KV cache to be moved between machines.

\yhedit{
Fetching KV cache from another machine causes a substantial delay, yet this network delay has not received sufficient attention.

\myparaq{Is it a new problem} 
Although some recent efforts also propose to send KV cache across GPUs to run multi-GPU inference, these systems assume that the KV cache is 
shared via high-speed links~\cite{zhong2024distserve, patel2023splitwise}, \eg direct NVLinks, which has bandwidth of up to several hundred Gbps. 
In these settings, the network delay to fetch KV cache can be negligible.
However, KV caches also need to be fetched over lower-bandwidth links, such as between regular cloud servers, where the bandwidth is usually in the single-digit Gbps range~\cite{skyplane}.
As illustrated in Figure~\ref{fig:motivate_baselines}b, in this setting, the delay of fetching KV cache into GPU memory can be as long as (or even longer than) prefill without the KV cache. 
}

\yhedit{

\mypara{Our approach}
This paper focuses on reducing the network delay in fetching the KV cache. 
To this end, we compress the KV cache by {\bf\em encoding} it into more compact bitstream representations (shown in Figure~\ref{fig:motivate_baselines}c).
This goal may seem similar to the recent works that drop words (tokens) from the text context or quantize the KV cache tensors~\cite{zhang2023h2o, scissorhand, hooper2024kvquant, kang2024gear, liu2024kivi}.
However, there is a key difference. 
These techniques reduce the {\bf\em run-time} GPU memory footprint of KV cache, thus retaining the tensor shapes of KV cache. 
In contrast, we reduce the {\bf\em transmission-time} size of KV cache by encoding it into compact bitstreams to reduce the network delay of sending it.
Moreover, there is a natural synergy---the KV cache shrunk by these recent works can still be encoded to further reduce the KV cache size and the network delay of sending KV caches.
}

\tightsection{\hspace{-0.0cm}\name: KV Cache Encoding and Streaming}
\label{sec:streamer}


The need to reduce KV cache transmission delay motivates a new module in LLM systems, which we call a {\em KV cache streamer}.
The KV cache streamer serves three roles: 
\begin{packeditemize}
\item {\em (1) Encoding} a given KV cache into more compact bitstream representations --- {\em KV bitstreams}. This can be done offline. 
\item {\em (2) Streaming} the encoded KV bitstream through a network connection of varying throughput.
\item {\em (3) Decoding} the received KV bitstream into the KV cache.
\end{packeditemize}

At first glance, our KV cache streamer may look similar to recent techniques (\eg~\cite{jiang2023longllmlingua,zhang2023h2o,scissorhand}) that compress long contexts by dropping less important tokens. Yet, they differ in crucial ways:

Those recent techniques aim at reducing the {\em run-time} size of the KV cache to accommodate the
GPU memory-size constraint or LLM input-window constraint, and yet we aim at reducing the {\em transmission-time} size of the KV cache to reduce network delay. 
As a result, previous techniques have to
maintain the KV caches'
shapes of large floating-point tensors so that the shrunk KV caches can be directly consumed by the LLM at the run-time;
meanwhile, they can use information during the generation phase to know which tokens in the context are more 
important to the particular query under processing. 
In contrast, 
 we need {\em not}
to maintain the original tensor shapes, and can encode them into more compact bitstreams and adapt their representation to network bandwidth. 
Meanwhile, we have to decide which compression scheme to use before a particular query is processed, and hence, we cannot use information
from the generation phase.

This paper presents {\bf \name}, a concrete design of the KV cache streamer.
First, \name uses a custom KV cache codec (encoder and decoder) to minimize the size of KV bitreams, by embracing several distributional properties of KV cache tensors (\S\ref{subsec:insight}).
This greatly reduces the bandwidth demand to transmit the KV cache, thus directly reducing TTFT. 
Second, when streaming the KV bitstreams under dynamic bandwidth, \name dynamically switches between different encoding levels or computing the KV cache on demand, 
in order to keep the TTFT within a given deadline while maintaining a high response quality.
The KV encoding/decoding incurs a negligible compute overhead and is pipelined with network transmission to minimize the impact on end-to-end delay.

\tightsection{\name Design}
\label{sec:design}

We now describe the design of \name, starting with the insights on KV cache (\S\ref{subsec:insight}) that inspires KV cache encoder (\S\ref{subsec:encoder}), followed by how \name adapts to bandwidth (\S\ref{subsec:controller}).

\tightsubsection{Empirical insights of KV cache}
\label{subsec:insight}

We highlight three observations on the characteristics of KV cache values. 
Though it is intrinsically hard to prove they apply to any LLM with any context, 
here, we use a representative workload to empirically demonstrate the prevalence of these observations.
The workload includes two LLMs of different capacities (Llama-7B and Llama-13B) and LongChat dataset~\cite{longchat2023} (which contains 100 long contexts between 9.2K and 9.6K tokens, randomly sampled from the whole set of 200 contexts), one of the largest datasets of long contexts. 
Details of this workload can be found in \S\ref{subsec:evalsetup}.


\begin{figure}[t]
    \centering
    \includegraphics[width=\linewidth]{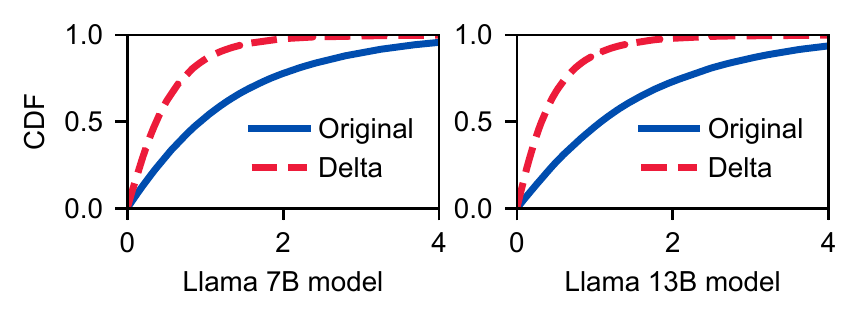}
    \vspace{-22pt}
    \tightcaption{Contrasting the distribution of the original values and the delta values. We model two Llama models with various long contexts (\S\ref{subsec:insight}). We show absolute values for clarity.} 
    \vspace{-10pt}
    \label{fig:locality}
\end{figure}

\subsubsection{{\bf \em Token-wise locality}}\label{subsubsec:locality}
The first observation is about how the K and V tensor values change {\em across tokens} in a context. Specifically, we observe that

\begin{insight}
    Within the same layer and channel, tokens in closer proximity have more similar K/V tensor values compared to tokens that are further apart.
\end{insight}

For each model, we contrast the distribution of K (or V) tensors' original values and the distribution of the {\em deltas}---the differences between K (or V) tensors' values at the same layer and channel between every pair of consecutive tokens in the contexts.
Figure~\ref{fig:locality} shows the distribution of absolute values in the original tensor and the deltas of one layer across all the contexts\footnote{We randomly sampled a single layer from the K tensor because the values in the different layers have different ranges.}.
In both models across the contexts, we can see that the deltas are much more concentrated around zero than the original values. 
Consequently, the variance of the deltas is 2.4-2.9$\times$ lower than that of the original values.
The token-wise locality of K and V tensors inspires \name to encode deltas rather than original values. 

This token-wise locality can be intuitively explained by the transformer's self-attention mechanism, which computes the KV tensors. 
The mechanism is mathematically equivalent to calculating the KV tensors of one token based on the KV tensors of the previous token.
This means KV tensors at one token are intrinsically correlated with those of the previous token.

\begin{figure}[t]
\centering
    \includegraphics[width=.99\linewidth]{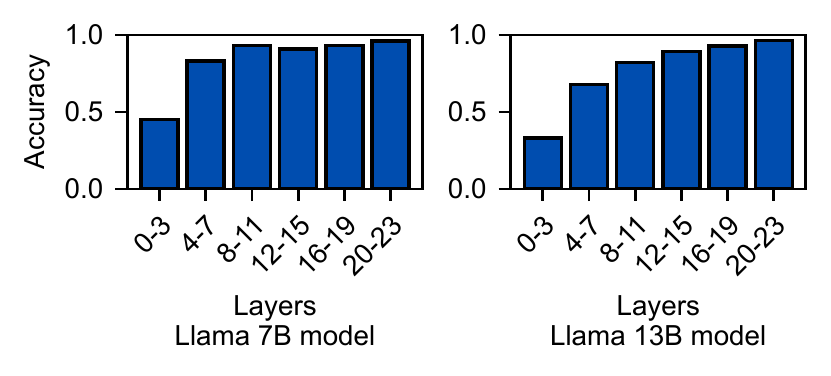}
    \vspace{-12pt}
    \tightcaption{Applying data loss to different layers of a KV cache has different impact on accuracy. (Same workload as Figure~\ref{fig:locality}).} 
    \label{fig:layer-sensitivity}
\end{figure}

\subsubsection{{\bf \em Layer-wise sensitivity to loss}}\label{subsubsec:sensitivity}
The second observation concerns how sensitive different values in the K and V tensors are to data loss. Our observation is the following: 

\begin{insight}
    The output quality of the LLM is more sensitive to losses in the KV cache values of the shallower layers than to losses in those of the deeper layers.
\end{insight}

The heterogeneous loss sensitivity on different layers suggests that our KV cache encoder should compress different layers differently.
Figure~\ref{fig:layer-sensitivity} shows how much accuracy is affected by applying data losses to the values of a specific layer group in the K and V tensors.
Here, we apply rounding as the data loss, and we compute the average resulting response accuracy (defined in \S\ref{subsec:evalsetup}) across 100 contexts in the dataset.
We can see that the average response accuracy drops significantly when the loss is applied to the early layers of a model while applying the same loss on the deeper layers has much less impact on the average response accuracy. This result holds consistently across different models we tested.

Intuitively, the deeper layers of a KV cache extract higher-level structures and knowledge than the shallower layers of a KV, which embed more primitive information~\cite{egeria, shao2024oneshot}.
As a result, the loss of information by removing precision on the early-layer cache might propagate and affect the later-layer cache, and thus hinder the model's ability to grasp the higher-level structures necessary to produce quality responses. 

\begin{figure}[t]
\centering
    \includegraphics[width=.99\linewidth]{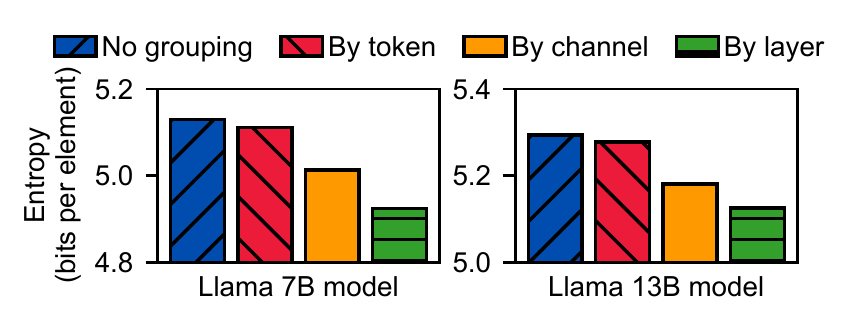}
    \vspace{-12pt}
    \tightcaption{Entropy (bits per element) when using different grouping strategies (Same workload as Figure~\ref{fig:locality}.)
    } 
    \label{fig:entropy}
\end{figure}

\subsubsection{{\bf \em Distribution along layers, channels, and tokens}}\label{subsubsec:entropy}
Finally, regarding the distributions of values along the three dimensions of KV cache---layers, channels, and token positions---we make the following observation.

\begin{insight}
Each value in a KV cache is indexed 
by its channel, layer, and token position. 
The information gain of grouping values by their channel and layer is significantly higher than the information gain of grouping values by their token position.
\end{insight}

Intuitively, this can be loosely interpreted as different KV values in the same channel (or layer) being more similar to each other than different KV values belonging to the same token position.
\yhedit{A possible explanation is that different channels or layers capture various features in the input~\cite{lin2021survey, devlin2018bert}. Some channels capture subject-object relationships, while others focus on adjectives. 
As for different layers, later layers capture more abstract semantic information than earlier ones according to prior works~\cite{lin2021survey, devlin2018bert}. 
On the other hand, within a given layer and channel, the KV values for different tokens are more similar, likely because of the self-attention mechanism, wherein each token's KV is derived from all preceding tokens.
We leave a more detailed examination to future work. 
}

To empirically verify the insight, we first group the values in the KV caches produced by the two models and 100 contexts based on their layers, channels, or token positions, and then compute the entropy of each group. 
Figure~\ref{fig:entropy} shows the average entropy (bits per element) when different grouping strategy is applied, including no grouping, grouping by tokens positions, grouping by channels, and grouping by layers. 
It shows grouping values by token positions reduces entropy much less than grouping by channel or layer. 

\subsection{KV cache encoding}
\label{subsec:encoder}

The aforementioned insights inspire the design of \name's KV cache encoder.
The encoding consists of three high-level steps (elaborated shortly):

First, it calculates the {\em delta tensors} (defined later) between the K and V tensors of nearby tokens. 
This is inspired by the token-wise locality observation (\S\ref{subsubsec:locality}) which suggests deltas between tokens might be easier to compress than the original values in the KV tensors.

Second, it applies different levels of quantization to different layers of the delta tensors.
The use of different quantizations at different layers is inspired by the observation of heterogeneous loss sensitivity (\S\ref{subsubsec:sensitivity}).

Third, it runs a lossless arithmetic coder to encode the quantized delta tensors into bitstreams.
Specifically, inspired by the observation in \S\ref{subsubsec:entropy}, the arithmetic coder compresses the values in each layer and channel separately (\S\ref{subsubsec:entropy}).

These steps may seem similar to video coding, which encodes pixels into bitstreams.
Video coding also computes the delta between nearby frames, quantizes them, and encodes the delta by arithmetic coding~\cite{cabac}.
Yet, blindly applying existing video codecs could not work well since they were only optimized for pixel values in natural video content. 
Instead, the exact design of \name is inspired by domain-specific insights on LLM-generated KV cache (\S\ref{subsec:insight}).

Next, we explain the details of each step. 

\begin{figure}[t]
\centering
     \includegraphics[width=.93\linewidth]{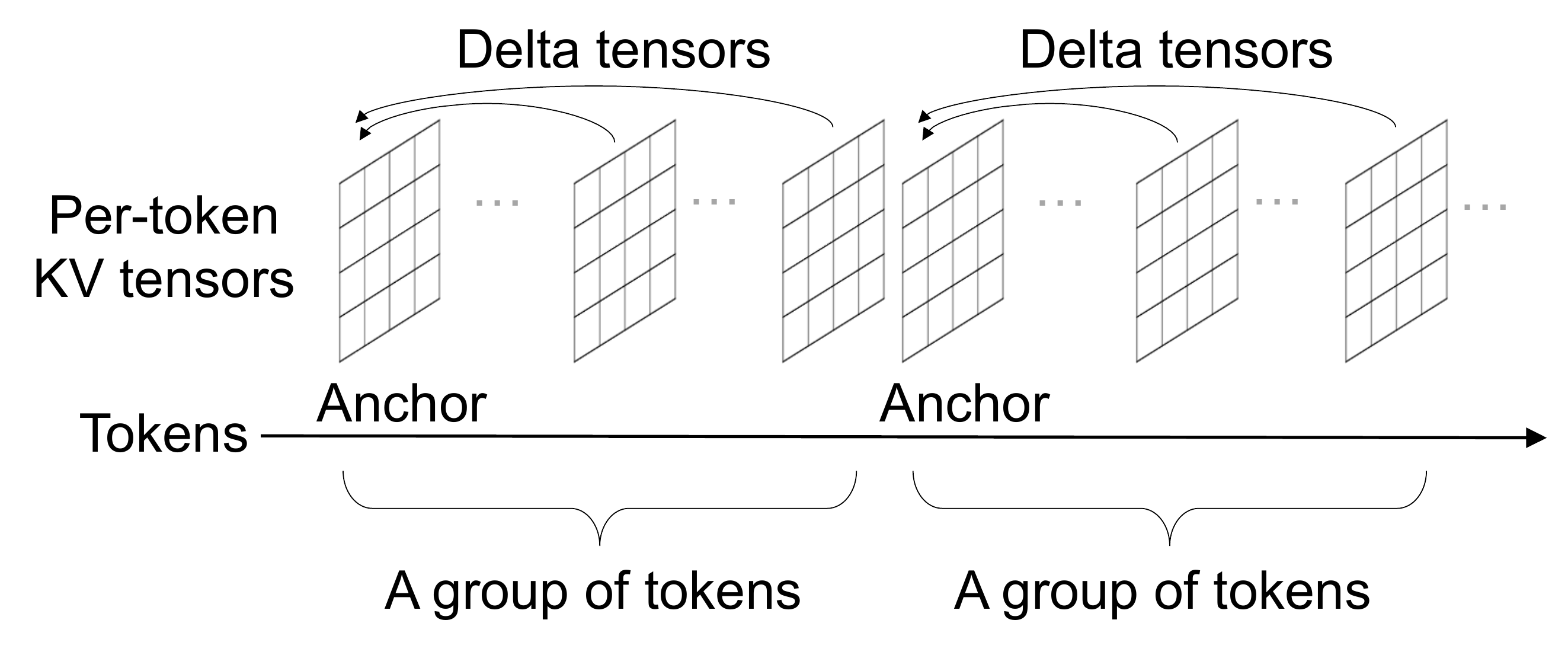}
    \tightcaption{Within a token group, \name computes delta tensors between  KV tensors of the anchor token and those of remaining tokens.}
    \vspace{-12pt}
    \label{fig:delta}
\end{figure}

\mypara{Change-based encoding} 
To leverage the token-wise locality, we first split the context into {\em groups of tokens} each containing ten contiguous tokens. 
As shown in Figure~\ref{fig:delta}, in each group, we independently (\ie without referencing other tokens) compress the KV tensor of the first token, called the {\em anchor token}, and then compress and record the {\em delta tensors} with respect to the anchor token for every other token.

This process is analogous to video coding, where the frames are separated into groups of {\em pictures}, within which it runs similar delta-based encoding. 
The difference, however, 
is that instead of compressing the delta between each pair of consecutive tokens, we reference the same anchor token for every token in the chunk. This allows us to do compression and decompression in parallel and saves time.

\mypara{Layer-wise quantization}
After partitioning the tokens into groups, \name uses quantization to reduce the precision of elements (floating points) in a KV cache so that they can be represented by fewer bits.
Quantization has been used recently to reduce attention matrices to pack longer contexts in GPU memory~\cite{flexgen}. However, in previous work, elements are uniformly quantized with the same number of bits without leveraging any unique properties of KV cache.
Driven by the insight of heterogeneous loss sensitivity (\S\ref{subsubsec:sensitivity}), we apply more conservative quantization (i.e., using more bits) on the delta tensors of earlier layers. 
Specifically, we split the transformer layers into three layer groups, the first (earliest) 1/3 of layers, the middle 1/3 of layers, and the last 1/3 of layers, and apply different amounts of quantization bin size on the delta tensors at each layer group respectively. 
The size of the quantization bin grows larger (\ie larger quantization errors)
from earlier to later layer groups. 
Following previous work~\cite{llmint8}, we use the \texttt{vectorwise} quantization method, which has been usually used for quantizing model weights. 

Note that we still use 8-bit quantization, a relatively high precision, on the KV cache of the anchor token (the first token of a token chunk). 
This is because these anchor tokens account for a small fraction of all tokens, but their precision affects the distribution of all delta tensors of the remaining tokens in a chunk.
Thus, it is important to preserve higher precision just for these anchor tokens.

\mypara{Arithmetic coding}
After quantizing the KV cache into discrete symbols, \name uses \textit{arithmetic coding}~\cite{ac} (AC) to losslessly compress the delta tensors and anchor tensors of a context into bitstreams.
Like other entropy coding schemes, AC assigns fewer bits to encode more frequent symbols and more bits to encode less frequent symbols.
For it to be efficient, AC needs {\em accurate, low-entropy} probability distributions of the elements in the KV cache.

Driven by the observation of the KV value distributions along layers, channels, and token positions (\S\ref{subsubsec:entropy}), we group KV values by channel and layer to obtain probability distributions.
Specifically, our KV encoder offline profiles a separate probability distribution for each channel-layer combination of delta tensors and another for anchor tensors produced by an LLM, and uses the same distributions for all KV caches produced by the same LLM.
\name uses modified AC library~\cite{mentzer2019practical} with CUDA to speed up encoding and decoding (\S\ref{subsec:impl}).
In \S\ref{subsec:eval:micro}, we empirically show that our method reduces the bitstream size by up to 53\% compared to the strawman of using one global symbol distribution.

\tightsubsection{KV cache streaming adaptation}

\begin{figure}[t]
\centering
    \includegraphics[width=0.95\linewidth]{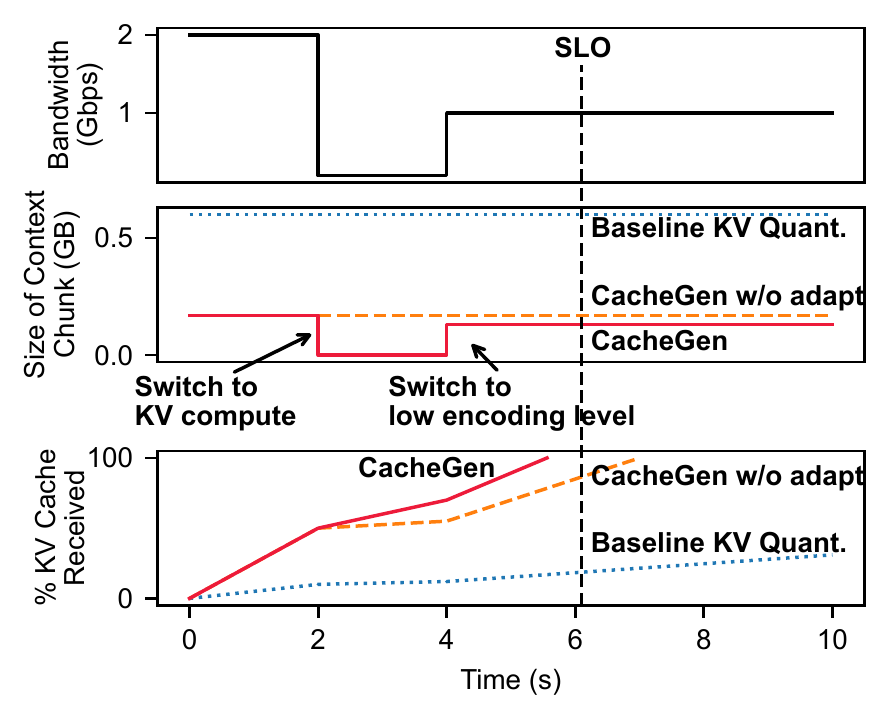}
    \vspace{-15pt}
    \tightcaption{Time Series demonstrating \name's adaptation logic under bandwidth variation.}
    \label{fig:adaptation-case}
    \vspace{-2pt}
\end{figure}

\label{subsec:controller}




Since the transmission of a KV cache may take up to hundreds of milliseconds to a few seconds, the available bandwidth may fluctuate during a transmission. 
Thus, streaming the encoded KV bitstreams at a fixed encoding level may violate a given service-level objective (SLO)~\cite{10.5555/3006357} of fetching the KV cache.\footnote{In practice, SLO is defined on TTFT. Once the KV cache of the long context is loaded in GPU, the remaining delay of one forward pass is marginal~\cite{vllm}.}
In Figure~\ref{fig:adaptation-case}, for example, 
at the start of the transmission, the available throughput 
is 2~Gbps, and if the bandwidth remains at 2~Gbps, sending a KV stream of 1~GB can meet the SLO of 4~seconds.
However, at $t=2s$, the throughput drops to 0.2~Gbps and only increases to 1~Gbps at $t=4s$, so the actual transmission delay increases from 4 seconds to 7 seconds, which violates the SLO.

\mypara{Workflow}
To handle variations in bandwidth, \name splits a context into multiple {\em context chunks} (or {\bf chunks} for short) of consecutive tokens and uses the KV cache encoder to encode each chunk into multiple bitstreams of different encoding (quantization) levels that can be decoded independently (explained shortly).
This can be done offline. 
When fetching a context, \name sends these chunks one by one, and each chunk can choose one of several {\em streaming configuration} (or {\bf configurations} for short): it can be sent at one of the encoding levels or can be sent in the text format to let the LLM recompute K and V tensors. 

\name adapts the configuration of each chunk while streaming the KV cache to keep the transmission delay within an SLO. 
Figure~\ref{fig:adaptation-case} illustrates an example adaptation where 
\name switches to sending text context and recomputing KV cache from the text at $t=2s$ due to the bandwidth drop, and at $t=4s$, since the bandwidth increases back to 1~Gbps, and \name switch to sending KV bitstreams of subsequent chunks at a smaller size. 
With our adaptation logic (specific algorithm in \S\ref{Adaptation Logic}), \name can meet the SLO.



However, to adapt efficiently, several questions remain.



First, {\em how to stream multiple chunks at different streaming configurations without affecting compression efficiency?}
To encode the chunks offline, \name first computes the KV cache of the entire context (\ie prefill) and splits the K and V tensors of the KV cache along the token dimension into sub-tensors, each of which contains the layers and channels of the tokens in the same chunk.
It then uses the KV encoder to encode the K or V sub-tensor of a chunk with different encoding (quantization) levels. 
Each chunk is encoded {\em independent} to other chunks {\em without} affecting the compression efficiency as long as a chunk is longer than a group of tokens.
This is because encoding the KV tensor of a token only depends on itself and its delta with the anchor token of the group of tokens (\S\ref{subsec:encoder}).
Thus, chunks sent with different encoding levels can be independently decoded and then concatenated to reconstruct the KV cache. 
In the case that a chunk is sent in text format, the LLM will compute its K and V tensors based on the previous chunk's KV tensors that have been received and decoded.\footnote{A similar concept has been used to split LLM input into prefill chunks for more efficient batching~\cite{agrawal2023sarathi}.}

{\em Would streaming chunks at different configurations affect generation quality?}
If one chunk is sent at a smaller-sized encoding level than other chunks (due to low bandwidth), it will have high compression loss on {\em that} single chunk, but this will not affect the compression loss of other chunks. 
That said, we acknowledge that if the bandwidth is too low to send most chunks at a high encoding level, the quality will still suffer.

Second, {\em how long should a context chunk be?}
We believe that the chunk length depends on two considerations. 
\begin{packedenumerate}
\item The encoded KV bitstream of a chunk size should not be too big because, otherwise, it cannot react to bandwidth changes in a timely manner.
\item The chunk should not be too small either since then we can not fully utilize the batching ability of GPU to compute KV tensors if text format is chosen.
\end{packedenumerate}
With these considerations in mind, we empirically pick 1.5K tokens as the default chunk length in our experiments\footnote{The chunk length is also long enough for the KV bitstream of each chunk to fill the sender's congestion window in our experiment setting.}, though more optimization may find better chunk lengths.

Thirdly, {\em how does \name decide the streaming configuration of the next chunk?} 
\yhedit{\name estimates the bandwidth by measuring the throughput of the previous chunk. It assumes this throughput will remain constant for the remaining chunks and calculates the expected delay for each streaming configuration accordingly. The expected delay is calculated by dividing its size by the throughput (more details in \S\ref{app:algorithm}). If there are bandwidth fluctuations, \name's reaction will be delayed by at most one chunk. Since one chunk is a small subset of the entire KV cache, this reaction is sufficiently fast to meet SLO (details in \S\ref{subsec:eval:adaption}).
}
%
It then picks the configuration that has the least compression loss (\ie text format or lowest encoding level) with an expected delay still within the SLO, and uses the configuration to send the next chunk.
For the first chunk, if some prior knowledge of the network throughput is available, \name will use it to choose the configuration of the first chunk the same way.
Otherwise, \name starts with a default medium encoding level (140~MB per chunk for Llama 7B, detailed setting in \S\ref{medium_encoding}).

\yhedit{Finally, \emph{how does \name handle the streaming of multiple requests? } 
When multiple requests arrive concurrently within \(T\) seconds, \name batches and streams them together. 
It can batch up to \(B\) requests, which is the maximum number that the GPU server can handle simultaneously. 
Each request is divided into chunks of the same size, even though the total number of chunks may differ among requests. 
For each chunk index \(c\), \name determines the number of requests \(N_c\) that include chunk \(c\). 
Using the throughput measured for the previous chunk \(c-1\), \name calculates the expected delays for each configuration by multiplying \(N_c\) by the delay for a single request. 
On the GPU servers, the requests are batched by padding their KV caches and processing them together.
}

\tightsection{Implementation}
\label{subsec:impl}
We implement \name with about 2K lines of code in Python, and about 1K lines of CUDA kernel code, based on PyTorch v2.0 and CUDA 12.0.

\mypara{Integration into LLM inference framework} 
\name operates the LLM through two interfaces: 
\begin{packeditemize}
    \item \texttt{calculate\_kv(context) -> KVCache}: given a piece of context, \name invokes LLM through this function to get the corresponding KV cache.
    \item \texttt{generate\_with\_kv(KVCache) -> text}: \name passes a KV cache to the LLM and lets it generate the tokens while skipping the prefilling of the context.
\end{packeditemize}
We implement these two interfaces in HuggingFace models using the \texttt{transformers} library~\cite{huggingface_transformer} with about 500 lines of Python code.
Both interfaces are implemented based on the \texttt{generate} function provided by the library.
For \texttt{calculate\_kv}, we let LLM only calculate the KV cache without generating new text, by passing the options of \texttt{max\_length} \texttt{= 0} and \texttt{return\_dict\_in\_generate} \texttt{=} \texttt{True} when getting the KV cache. 
The \texttt{generate\_with\_kv} is implemented by simply passing the KV cache via the \texttt{past\_key\_values} argument when calling the \texttt{generate} function.
Similar integrations are also applicable to other LLM libraries, such as FastChat~\cite{zheng2023judging}, llama.cpp~\cite{llamacpp}, and GGML~\cite{ggml}.

We have also integrated \name in LangChain~\cite{langchain}, a popular LLM application framework.
\name is activated in the \texttt{\_generate} function of  LangChain's  \texttt{BaseLLM} module. 
\name first checks whether the KV cache of the current context already exists (explained shortly).
If so, \name invokes \texttt{generate\_with\_kv} to start generating new texts.
Otherwise, \name will invoke \texttt{calculate\_kv} to create the KV cache first before generating new texts.

\mypara{KV cache management in \name} To manage the KV cache, \name implements two modules:
\begin{packeditemize}
    \item \texttt{store\_kv(LLM) --> \{chunk\_id: encoded\_KV\}}: calls
    \\ \noindent 
    \texttt{caculate\_kv}, splitting the returned KV cache into context chunks, and encodes each chunk. 
    Then, it stores a dictionary on the storage server, where it maps the \texttt{chunk\_id} to the encoded bitstreams for the K and V tensors for the corresponding chunk.
    \item \texttt{get\_kv(chunk\_id) --> encoded\_KV} fetches the encoded KV tensors corresponding to \texttt{chunk\_id} on the storage server and transmits it to the inference server. 
\end{packeditemize}
Whenever a new piece of context comes in, \name first calls \texttt{store\_kv}, which first generates the KV cache, and then stores the encoded bitstreams on the storage server. 
At run time, \name calls \texttt{get\_kv} to fetch the corresponding chunk of KV cache and feed into \texttt{generate\_with\_kv}. 


\mypara{Speed optimization for \name} 
To speed up the encoding and decoding of KV cache, we implemented a GPU-based AC library~\cite{mentzer2019practical} with CUDA to speed up encoding and decoding. 
Specifically, each CUDA thread is responsible for encoding/decoding the KV cache from the bitstream of one token. 
The probability distributions are obtained by
counting the frequencies of quantized symbols in the KV feature for the corresponding context.
We also pipeline the transmission of context chunk $i$ with the decoding of context chunk $i-1$. 




\tightsection{Evaluation}
\label{sec:eval}


The key takeaways of our evaluation are: 
\begin{packeditemize}
    \item Across four datasets and three models, \name can reduce TTFT (including both network and compute delay) by 3.1-4.7$\times$ compared to prefill from text context,
    and by 3.2-3.7$\times$ compared to the quantization baseline (\S\ref{subsec:overall_eval}). 
    \item \name's KV encoder reduces the bandwidth for transferring KV cache by 3.5-4.3$\times$ compared to the quantization baseline (\S\ref{subsec:overall_eval}). 
    \item \name's reduction in bandwidth usage is still effective when applied to recent context compression baselines~\cite{zhang2023h2o, jiang2023longllmlingua}. 
    \name 
    further reduces the bandwidth usage by 3.3-4.2$\times$, compared to applying quantization on context compression baselines (\S\ref{subsec:overall_eval}). 
    \item \name's improvement is significant across various workloads, including different context lengths, network bandwidths, and numbers of concurrent requests (\S\ref{subsec:eval:sensitivity}). 
    \item \name's decoding overhead is minimal, in delay and compute, compared with LLM inference itself (\S\ref{subsec:eval:micro}).
\end{packeditemize}

\tightsubsection{Setup}
\label{subsec:evalsetup}

\begin{table}[t]
\resizebox{0.8\columnwidth}{!}{
  \footnotesize
  \begin{tabular}{ccccc}
    \toprule
    \textbf{Dataset}  & \textbf{Size}&  \textbf{Med.}  & \textbf{Std.}  & \textbf{P95}\\
    \hline
    LongChat~\cite{longchat2023}  &200  & 9.4K    & 164  & 9.6K \\ 
    TriviaQA~\cite{joshi2017triviaqa} &200  &  9.3K  &  4497   &  15K \\
    NarrativeQA~\cite{kočiský2017narrativeqa} &200  & 14K     & 1916    & 15K \\
    WikiText~\cite{wikitext}  &62 &  5.9K  & 4548  & 14.8K \\
    \bottomrule
  \end{tabular}
  }
  \vspace{0.2cm}
  \tightcaption{Size and context lengths of datasets in the evaluation. }
  \label{tab:datasets}
  \vspace{-24pt}
\end{table}

\begin{figure*}[t]
\centering
\includegraphics[width=0.65\linewidth]{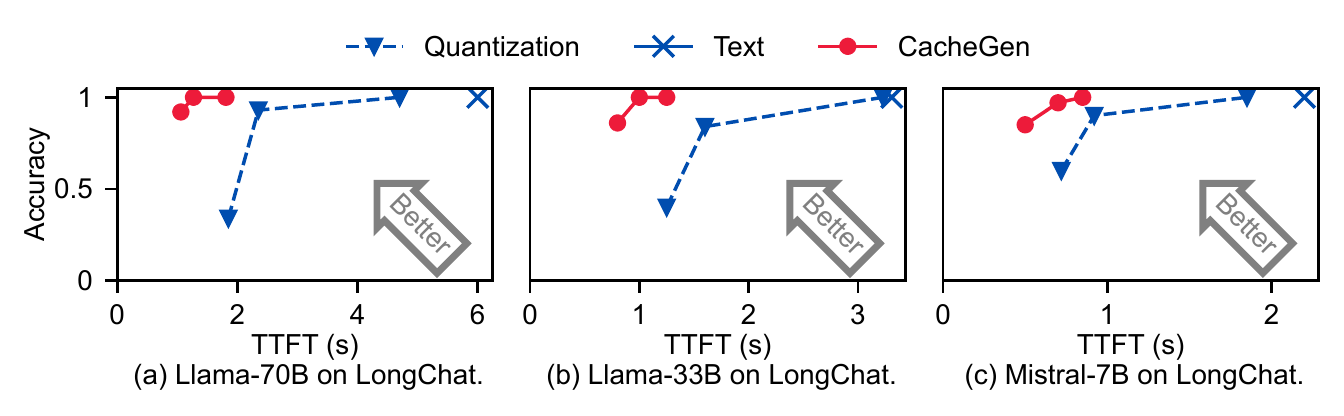}
\vspace{-10pt}
\includegraphics[width=0.65\linewidth]{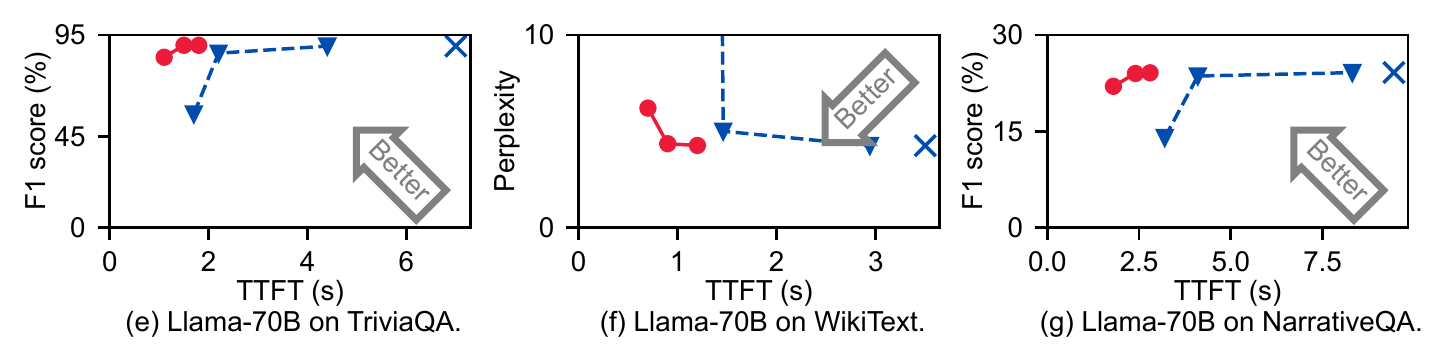}

\tightcaption{{\bf Time-to-first-token (TTFT):} Across different models and different datasets, \name reduces TTFT with little negative impacts on quality (in accuracy, perplexity or F1 score). 
}
\label{fig:e2e-longchat}
\end{figure*}











\mypara{Models} We evaluate \name on three models of different sizes, specifically the fine-tuned versions of Mistral-7B, Llama-34B, and Llama-70B. 
All models are fine-tuned such that they can take long contexts (up to 32K). 
We did not test \name on other LLMs (\eg OPT, BLOOM) because there are no public fine-tuned versions for long contexts to our best knowledge.

\mypara{Datasets}
We evaluate \name on 662 contexts from four different datasets with different tasks (Table~\ref{tab:datasets}): 
\begin{packeditemize}
    \item {\em LongChat:} The task is recently released~\cite{longchat2023} to test LLMs on queries like ``What was the first topic we discussed?'' by using all the previous conversations as the context.  
    Most contexts are around 9.2-9.6K tokens. 
    \item {\em TriviaQA:} The task tests the reading comprehension ability of the LLMs~\cite{bai2023longbench}, by giving the LLMs a single document (context), and letting it answer questions based on it. The dataset is part of the LongBench benchmark~\cite{bai2023longbench} suite. 
    \item {\em NarrativeQA:} The task is used to let LLMs answer questions based on stories or scripts, provided as a single document (context).
    The dataset is also part of LongBench.
    \item {\em Wikitext:} The task is to predict the probability of the next token in a sequence based on the context consisting of relevant documents that belong to a specific Wiki page~\cite{wikitext}.
\end{packeditemize}

\yhedit{
The dataset we used to design \name's encoder is a subset of 
the datasets we used to evaluate \name.
This is for showing the insights in \S\ref{subsec:insight} are generalizable to different datasets. 
}




\mypara{Quality metrics}
We measure generation quality using the standard metric of each dataset.
\begin{packeditemize}


\item {\em Accuracy} is used to evaluate the model's output on the LongChat dataset. 
The task predicts the first topic in the conversational history between the user and the LLM. 
The accuracy is defined as the percentage of generated answers that exactly includes the ground-truth topic. 

\item {\em F1 score} is used to evaluate the model's response in the TriviaQA and NarrativeQA datasets.
It measures the probability that the generated answer matches the ground-truth answer of the question-answering task.
\item {\em Perplexity} is used to evaluate the model's performance on the Wikitext dataset. 
The perplexity is defined as the exponentiated average negative log-likelihood of the next token~\cite{ppl1, ppl2}.
A low perplexity means that the model likely generates the next token correctly.
While perplexity does not equate to text-generation quality, it is widely used as a proxy~\cite{adiwardana2020humanlike} 
to test the impact of pruning or quantizing LLMs on generation performance~\cite{llmint8,smoothquant,liu2023scissorhands, roy2021efficient}.
\end{packeditemize}

\yhedit{\mypara{System metrics} We compare \name with baselines with two system-wise metrics.

\begin{packeditemize}
    \item {\em Size of KV cache} is the size of the KV cache after compression, this measures
    the bandwidth needed to load KV caches.
    \item {\em Time-to-first-token (TTFT)} is the time from the arrival of the user query to the 
    generation of the first token. This includes the loading delay of the KV cache 
    and the prefill delay of the new questions.
    This is a metric widely used in industry~\cite{anyscale_2023,anyscale_2023_reproducible,databricks_2023_llm_inference}
    and recent works~\cite{promptcache, liu2024andes}. 
\end{packeditemize}

}


\mypara{Baselines} 
We compare \name with baselines that do not change the contexts or model (more baselines in \S\ref{subsec:eval:micro}). 
\begin{itemize}
\item {\em ``Default quantization''} uses the uniform quantization of KV cache,
specifically the same quantization level \yhedit{(\ie 3, 4, 8 bits)} for every layer in the LLM 
(which was used in~\cite{flexgen}).
\item {\em ``Text context''} fetches the text of the context and feeds it to LLM to generate the KV cache for it. 
It represents the design of minimizing data transmission but at the expense of high computation overhead. 
We use the state-of-the-art inference engine, vLLM~\cite{vllm},  to run the experiments. 
vLLM's implementation already uses xFormers~\cite{xFormers2022}, which includes speed and memory-optimized Transformers CUDA kernels and has shown much faster prefill delay than HuggingFace Transformers.   
This is a very competitive baseline. 
\item {\em ``Context compression''} either drops tokens in the text context (LLMlingua~\cite{jiang2023longllmlingua}) or in the KV cache (H2O~\cite{zhang2023h2o}). 

\end{itemize}

\begin{figure*}[t]
\centering
\includegraphics[width=0.96\linewidth]{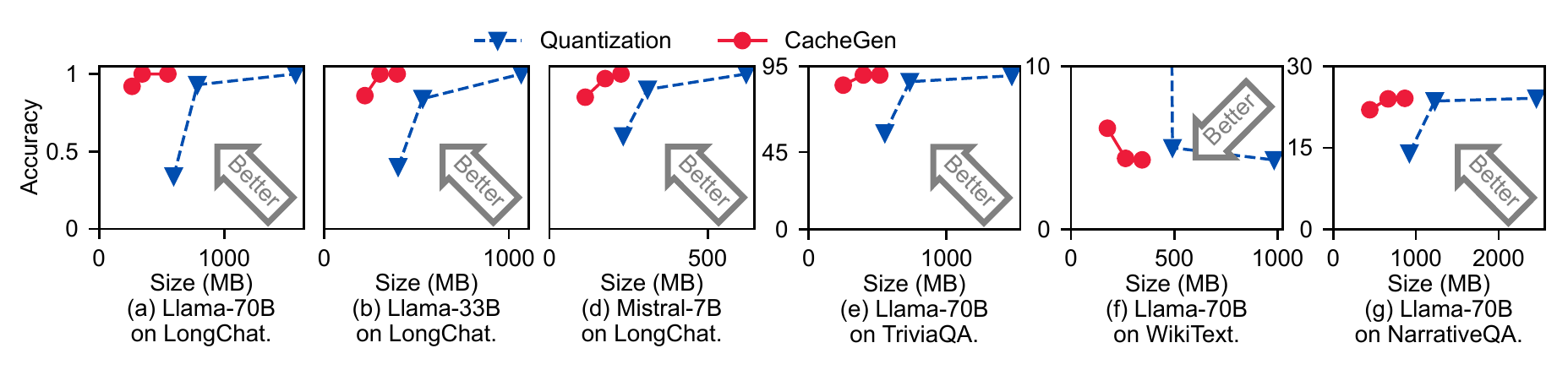}
\vspace{-10pt}

\tightcaption{{\bf Reducing KV cache size:} Across various models, \name reduces size of KV cache with little accuracy decrease on various datasets. 
}
\vspace{-5pt}
\label{fig:e2e-longchat-quant}
\end{figure*}

\mypara{Hardware settings} We use an NVIDIA A40 GPU server with four GPUs to benchmark our results. 
The server is equipped with 384GB of memory and two Intel(R) Xeon(R) Gold 6130 CPUs with Hyper-threading and Turbo Boost enabled by default.


%

\extightsubsection{Overall improvement}
\label{subsec:overall_eval}

We first show the improvement of \name over the baselines, as described in \S\ref{subsec:evalsetup}. 

\mypara{TTFT reduction} 
Figure~\ref{fig:e2e-longchat} demonstrate \name's ability to reduce TTFT, across three models and four datasets. Under bandwidth of 3~Gbps, 
compared to text context, \name is able to reduce TTFT by 3.1-4.7$\times$. 
Compared to default quantization, \name is able to reduce TTFT by 3.2-3.7$\times$.



It is important to note that even compared with 8-bit quantization, an almost lossless KV cache compression technique across the four datasets, \name can still reduce the TTFT by 1.67-1.81$\times$. 
\name's reduction in TTFT is a result of a shorter transmission delay to send the smaller KV caches.


\mypara{Reduction on KV cache size}
Figure~\ref{fig:e2e-longchat} show that, across four datasets and three models, 
\name's KV encoder reduces the KV cache size by 3.5-4.3$\times$ compared to default quantization when achieving similar performance for downstream tasks after decoding. 
Thus, it achieves better quality-size trade-offs across different settings.
The degradation caused by lossy compression is marginal---the degradation is no more than 2\% in accuracy, less than 0.1\% in F1 score, and less than 0.1 in perplexity~\cite{huggingface_ppl}.

Some example text outputs for different baselines are available in \S\ref{app:output}.

\mypara{Gains over context compression baselines}
\yhedit{We also apply \name to further reduce the size of  context compression baselines' KV cache, including H2O and LLMlingua.
Note that H2O drops tokens from KV cache which have low attention scores. 
Specifically, it requires the query tensors of the prompt to compute the attention scores in order to determine which tokens to drop. 
The query tensors of the prompts are not present in the offline compression stage. 
In our experiments,
we implement an \emph{idealized} version of H2O, where the query tensors of the prompts are used in the offline compression stage.


\begin{figure*}[]
\centering
\includegraphics[width=\linewidth]{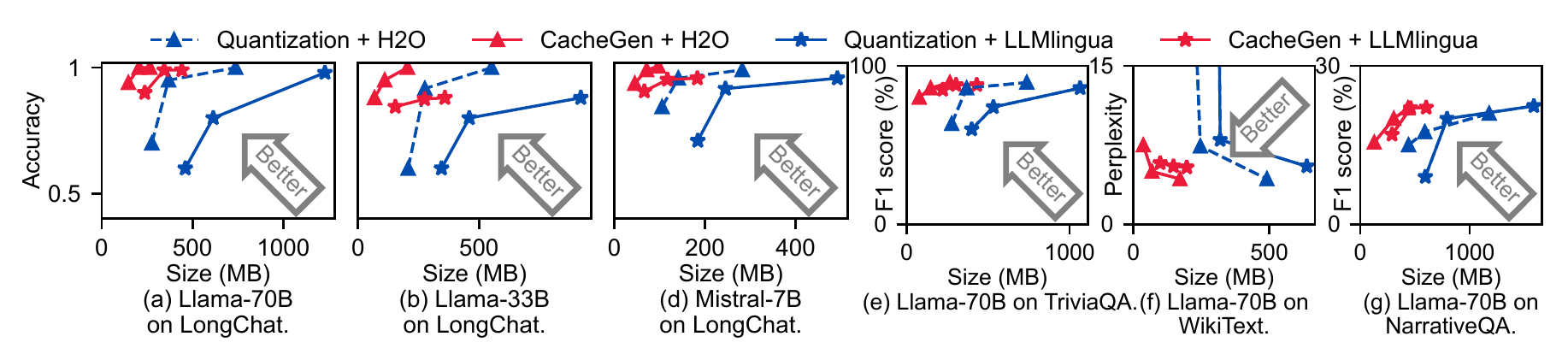}
\vspace{-20pt}

\tightcaption{
\yhedit{{\bf Reducing KV cache size {\em on top of} H2O~\cite{zhang2023h2o} and LLMlingu~\cite{jiang2023longllmlingua}:} Across different models, \name further the size of KV cache, compared to the KV cache shortened by H2O, with little accuracy decrease on different datasets. }
}
\label{fig:e2e-longchat-h2o}
\end{figure*}




As shown in Figure~\ref{fig:e2e-longchat-h2o}, compared to the context compression baseline, H2O~\cite{zhang2023h2o}, \name can further reduce compressed KV cache (in floating point).
Specifically, 
\name reduces the size of KV cache by 3.5--4$\times$ compared to the H2O's quantized KV caches, and 3.3--4.2$\times$ compared to LLMlingu's quantized KV caches,
without losing quality.
This suggests that even after condensing contexts by H2O and LLMlingua, the resulting KV caches may still have the statistical observations behind \name's KV encoder.
Thus, the techniques used in \name's encoder remain beneficial when we encode the KV cache after applying these techniques.}


\mypara{Understanding \name's improvements}
\name outperforms various baselines for slightly different reasons. 
Compared to the text context baseline, \name has lower TTFT, because it reuses KV cache to avoid the long prefill delay for processing long contexts. 
Compared to the basic quantization baseline, \name compresses KV cache with layer-wise dynamic quantization and further encodes the KV cache tensors into bitstreams, thus able to reduce the transmission delay. 

Finally, compared to H2O and LLMlingua, two recent context-condensing techniques, \name can still compress the KV cache produced by H2O.
In short, H2O and other context-condensing techniques all prune contexts at the token level and their resulting KV caches are in the form of floating-point tensors, so \name is complementary and can be used to further compress the KV cache into much more compact bitstreams.

\begin{figure}[t]
    \centering
    \includegraphics[width=.9\linewidth]{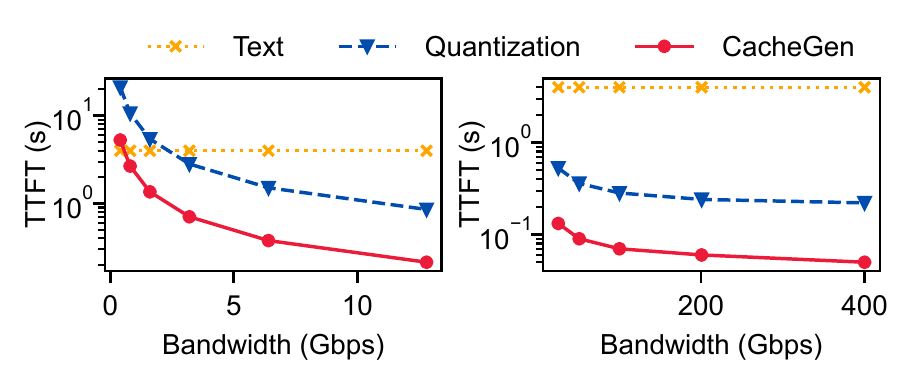}
    \vspace{-15pt}
    \tightcaption{\name improves TTFT under a wide range of different bandwidths. Plotted with Mistral-7B.
    y-axis is log scale. }
    \vspace{-0.2cm}
    \label{fig:sensitivity-bw}
\end{figure}

\tightsubsection{Sensitivity analysis}
\label{subsec:eval:sensitivity}

\mypara{Available bandwidth} \yhedit{The left and right figures in Figure~\ref{fig:sensitivity-bw} 
compare the TTFT of \name with baselines under a wide range of bandwidth from 0.4--15 Gbps
 and 15--400 Gbps, while we fix the context length at 16K tokens. 
 We can see that \name consistently outperforms baselines under almost all bandwidth situations. 
Arguably, the \emph{absolute reduction} in TTFT becomes smaller under high bandwidth (over 20Gbps), compared to 
the quantization baseline, since both the quantization baseline and \name can 
transfer KV caches much faster. 
 }


\begin{figure}[t]
    \centering
    \includegraphics[width=0.9\linewidth]{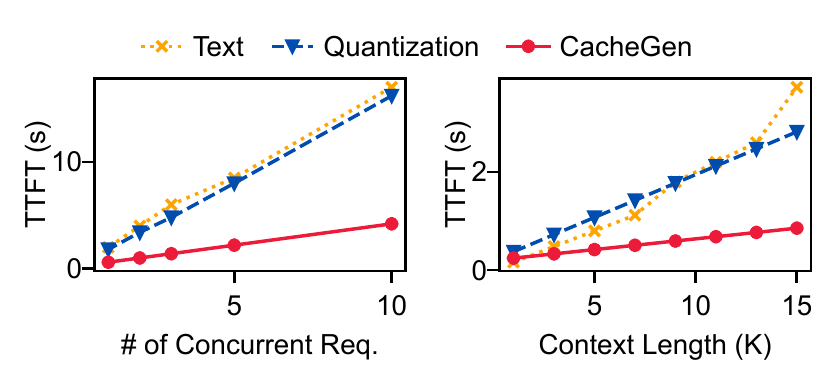}
    \vspace{-10pt}
    \tightcaption{\name consistently reduces TTFT when there are multiple concurrent requests on one GPU. Plotted with Mistral-7B. }
    \label{fig:sensitivity-gpu}
    \vspace{-10pt}
\end{figure}

\mypara{Number of concurrent requests}
The left side of Figure~\ref{fig:sensitivity-gpu} shows the TTFT under different numbers of concurrent requests.
When the number of concurrent requests increases (\ie fewer available GPU cycles for one individual query), \name significantly reduces TTFT than the baselines.
This is because the amount of computation required for prefilling on a long input (9.6K in this case) is huge, as discussed in \S\ref{subsec:context}. 
\S\ref{app:space} shows \name's improvement over a complete space of workloads of different bandwidth and GPU resources.



\mypara{Context lengths}
The right side of Figure~\ref{fig:sensitivity-gpu}  compares \name's TTFT with the baselines under different input lengths from 0.1K to 15K tokens under a fixed network bandwidth of 3~Gbps.
When the context is long, the gain of \name mainly comes from reducing the KV cache sizes.
And when the context is short (below 1K), \name will automatically revert to loading the text context as that yields a lower TTFT.




\subsection{KV streamer adaptation}
\label{subsec:eval:adaption}

\begin{figure}[t]
    \centering
    \includegraphics[width=0.9\linewidth]{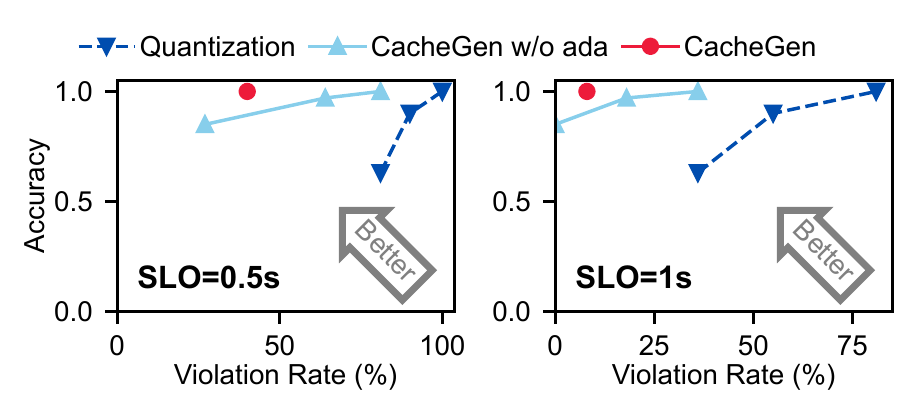}
    \vspace{-0.2cm}
    \tightcaption{
    \name reduces SLO violation rate over \name without adaptation and the quantization baseline. Plotted with Mistral-7B model. }
    \vspace{-5pt}
\label{fig:adaptation_res}
\end{figure}

The adaptation logic described in \S\ref{subsec:controller} allows \name to adapt to bandwidth changes and achieve good quality while meeting the SLO on TTFT. 
In Figure~\ref{fig:adaptation_res}, we generate bandwidth traces where each context chunk's bandwidth is sampled from a random distribution of 0.1 -- 10~Gbps.
Each point is averaged across 20 bandwidth traces on the LongChat dataset.
We can see that \name significantly outperforms the quantization baseline and \name without adaptation. 
Specifically, given an SLO on the TTFT of 0.5s, \name reaches the same quality as the quantization baseline with a 60\% lower SLO violation rate. 
Under an SLO of 1s, 
\name reaches the same quality as the quantization baseline, while reducing the SLO violation rate from 81\% to 8\%. 
The reason why \name has a lower SLO violation rate is that
when the bandwidth drops, \name can dynamically reduce the quantization level or fall back to the configuration of computing text from scratch, while the quantization baseline and \name without adaptation cannot. 



\tightsubsection{Overheads and microbenchmarks} 
\label{subsec:eval:micro}

\begin{figure}[t]
\centering
    \includegraphics[width=0.9\linewidth]{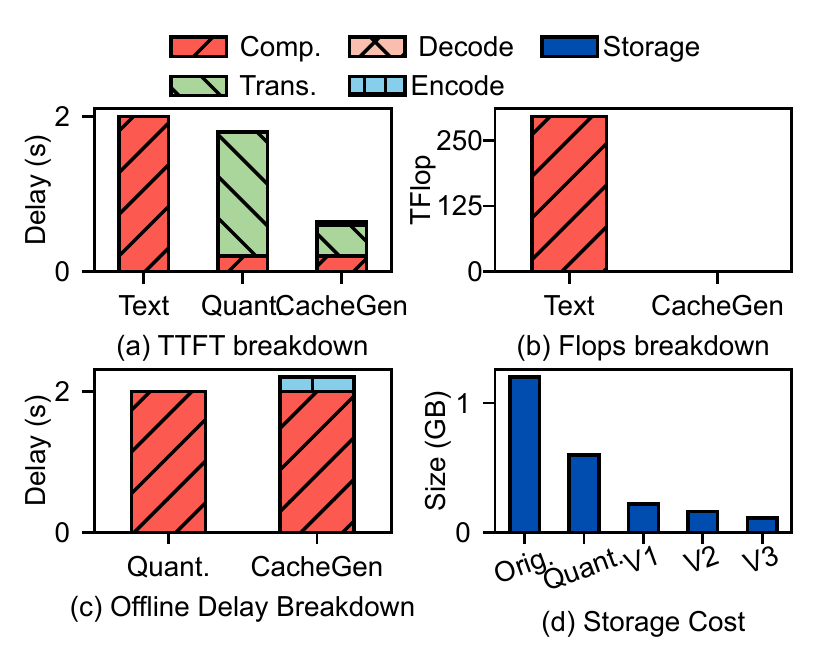}
    \vminten
    \tightcaption{(a) The breakdown of TTFT for text context, quantization baseline, and \name.  
    (b) Computation overhead of the text baseline and \name. 
    (c) Offline delay breakdown for baseline quantization and \name. 
    (d) The storage cost for \name, quantization baseline and the uncompressed KV cache. Plotted with Mistral-7B.
}
    \label{fig:overhead}
\end{figure}

\mypara{Decoding overhead}
While having a better size-quality and TTFT-quality trade-off, \name requires an extra decoding (decompression) step compared to the quantization baseline.
\name minimizes the decoding overhead by accelerating it with GPU-based implementation and pipelining the decoding of context chunks with the transmission of the context chunks, so as shown in Figure~\ref{fig:overhead}a, the decoding has minimal impact on the end-to-end delay.
It is also important to note that although \name's decoding is performed on GPU (see \S\ref{subsec:impl}), the amount of computation needed by \name's decoding module is negligible compared to the baseline that generates KV cache from text context. 


\mypara{Offline encoding and storage overheads}
\label{subsec:eval:storage}
Unlike prior methods that compress each context only once, \name compresses it into
multiple versions (\S\ref{subsec:controller}). 
\name compresses each context almost as fast as the baselines because the encoding delay is very small (200 ms), as shown in Figure~\ref{fig:overhead}c.
Figure \ref{fig:overhead}d evaluates the overhead in storage.
We can see that despite needing to encode and store multiple bitstream representations, the total
storage cost for \name is on par with the quantization baseline. 

\begin{figure}
\centering
        \includegraphics[width=0.8\linewidth]{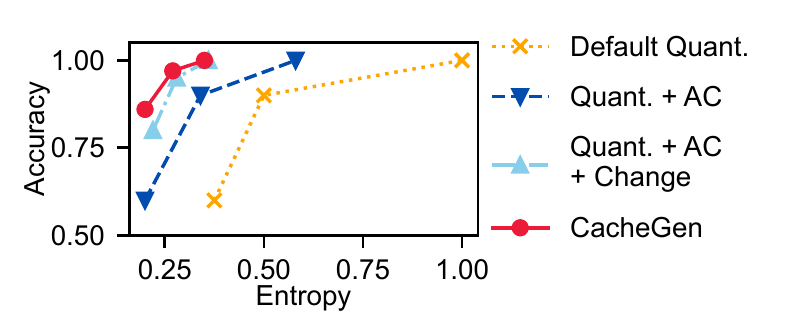}
        \vspace{-10pt}
        \tightcaption{Contributions of individual ideas behind KV encoder: change-based encoding, layer-wise quantization, and AC based on channel-layer grouping.
        }
        \vspace{-5pt}
        \label{fig:insights}
\end{figure}

\begin{figure}[t]
\centering
    \includegraphics{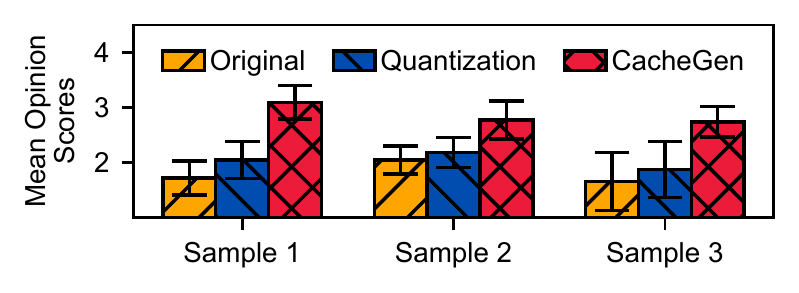}
    \vspace{-10pt}
    \tightcaption{Real user study shows \name improves QoE significantly over other baselines. }
    \label{fig:user-study}
    \vspace{-5pt}
\end{figure}

\mypara{Ablation Study}
To study the impact of individual components in \name's KV encoder,  Figure~\ref{fig:insights} progressively adds each idea into the baseline of uniform quantization and default AC, starting with the use of our AC that uses probability distribution for each channel-layer combination, then change-based encoding, and finally layer-wise quantization.
\yhedit{
As shown in the figure, \name's AC and change-based encoding significantly improve upon the uniform quantization. This indicates that removing the constraint of maintaining the tensor format of KV cache, and encoding them into bitstreams with our change-based encoding and AC can further reduce the size of KV cache after quantization.}

\mypara{Quality of Experience}
We performed an IRB-approved user study to validate the effectiveness of \name. We selected three conversation histories from the LongChat dataset used in previous evaluations. For each user, we first present the conversation history with ChatGPT. Then we show the same response but produced by different pipelines by adding different TTFTs and letting users rate the quality of response. With 270 ratings collected from Amazon MTurk~\cite{mturk}, we show that \name consistently outperforms other pipelines in QoE with shorter TTFT in Figure~\ref{fig:user-study}.

Evaluation results of \name with more baselines are available \S\ref{appendix:intrusive}, including using a smaller-sized model to speed up TTFT and Gisting, another context-shrinking technique.

\vspace{-0.2cm}
\section{Related Work}
\label{sec:related}

\mypara{Faster LLM serving}Most LLM systems research aims to speed up LLM training~\cite{deepspeed, shoeybi2019megatron} or make serving systems faster. \name aims at speeding up LLM serving systems by focusing on TTFT reduction. Others explore approximately parallelizing generation~\cite{Leviathan2022FastIF, miao2023specinfer}, accelerating inference on edge devices~\cite{yi2023edgemoe}, quantizing LLM weights~\cite{aminabadi2022deepspeed}, reducing memory I/O of GPU on-chip SRAM~\cite{dao2022flashattention} and reducing self-attention computation complexity~\cite{roy2021efficient}, better scheduling strategies~\cite{wu2023fast,orca, zhong2024distserve, patel2023splitwise, agrawal2023sarathi}, and GPU memory utilization~\cite{vllm}. 
Another line of work optimizes the communication delay of transmitting KV cache between GPUs, either by smart model parallelism strategies~\cite{zhong2024distserve, patel2023splitwise} or by implementing a new attention operation~\cite{lin2024infinitellm}. This operation transmits query vectors to the GPUs that host smaller blocks of KV cache during the decoding phase.
A common approach for faster inference without modifying the LLMs is by {\em caching the KV} of previously used inputs for {\em one} LLM query~\cite{flexgen, zhang2023ho, scissorhand, pope2022efficiently, miao2023specinfer, wolf-etal-2020-transformers}. 
\name works as a module to enable reuse of KV caches {\em  across  multiple} LLM queries in these frameworks~\cite{vllm,agrawal2023sarathi,promptcache,retro}.


\mypara{Longer LLM contexts} Recent efforts aim at enabling LLMs to process very long contexts~\cite{xu2024retrieval}.
The challenge is to fit the large attention matrices of longer contexts into limited GPU memory. 
This is enabled by offloading parts of the attention matrices~\cite{flexgen},
using external knowledge via KNN~\cite{wu2022memorizing}, approximating via 
retraining self-attention to only attend to top-k keys~\cite{bertsch2023unlimiformer,ainslie2020etc}, 
mapping long inputs to smaller latent spaces~\cite{pmlr-v162-hawthorne22a} and using local windowed, dilated or sparse~\cite{ding2023longnet,beltagy2020longformer,zaheer2021big} attention to scale to inputs of $\sim$1 billion tokens.
Longer contexts inflate the KV cache and \name aims to address this by fast remote loading of the KV cache.

\mypara{Context shortening} Efforts on shortening long contexts relate well to \name. They aim to select the most important text segments and prune the rest. Using similarity between the user query and the relevant documents~\cite{retro}, only keeping tokens that are less attended to by the prompt (\ie heavy-hitter tokens) ~\cite{scissorhand, zhang2023ho} or by hybrid policies including keeping nearby tokens or heavy-hitter tokens~\cite{ge2023model}, using query-aware compression with document reordering to reduce loss-in-the-middle~\cite{jiang2023longllmlingua, ribar2023sparq} have been explored. 
All these methods need to know the query, else they risk dropping potentially important tokens and they keep the KV cache intact, to fit into limited GPU memory. Some works retrain LLM models to use contexts rewritten by gisting~\cite{gisting} or auto-encoding~\cite{autoencoder-incontext}.

\name differs by compressing the KV cache into bitstreams instead of shortening the context. \name's KV compression does not need to know the query/prompt and doesn't risk quality loss from dropping potentially important tokens. It allows for better compression rates by leveraging distributional properties of KV caches and achieves better delay-quality trade-offs than existing context compressors (\S\ref{subsec:eval:micro}). \name also does not need to retrain the LLM.

\mypara{Tensor compression}\name's KV cache encoding is essentially a tensor compression technique tailored for LLM's. General tensor compression has been intensively studied~\cite{zhao2016tensor,tensor_decomp}. In DNN training, tensor compression has been used to compress gradient updates of DNN weights (\eg~\cite{espresso,MLSYS2022_773862fc,agarwal2020accordion}).
KV caches and gradients have very different properties.
DNN training systems often leverage the sparsity of gradients which occurs due to methods like~\cite{chen2023sparse, zehui2019dropattention, child2019generating}. However the KV cache is not known to be sparse in general.

\mypara{Retrieval augmented generation(RAG)}RAG~\cite{retro, atlas, fid,rubin2023long,lewis2020retrieval, wang2023zemi, ram2023incontext} focuses on retrieving relevant documents to the query via vector based~\cite{chen2017reading, Yang_2019, nie2019revealing} or DNN-based~\cite{lewis2020retrieval, karpukhin2020dense,xiong-etal-2021-progressively, reranking} similarity search algorithms and feeding it as context to generate the answer. We envision RAG as a fitting use case for \name. Many LLM inference platforms support feeding KV caches as retrieved context instead of text~\cite{hf2020, Chase_LangChain_2022}. Some works have also attempted to define a systematic way to choose which KV cache to reuse\cite{gim2023prompt}. Another approach is to have LLM applications that cache the query's generated answers to reduce repetitive query costs~\cite{ZillizTech2023, Martinez2023}. While caching answers is useful for reuse, \name provides a more generic way to incorporate context reuse and can generate better-quality answers.




\section{Discussion and Limitations}

\yhedit{\mypara{Compatibility with other KV-cache compression work} Emerging techniques like smart quantization~\cite{kang2024gear, liu2024kivi, hooper2024kvquant} are {\em complementary} with \name. 
After quantization, \name can still apply delta encoding and arithmetic coding, as shown in Figure~\ref{fig:e2e-longchat-h2o}.

\mypara{Incremental KV cache streaming} Future work includes extending \name to stream KV caches incrementally, akin to Scalable Video Coding (SVC)~\cite{svc}, by initially sending low-quality KV caches and then incrementally improving quality by sending differences.

\mypara{Context reuse in real-world LLM applications} In \S\ref{subsec:context}, we explain why contexts are likely reused across requests using anecdotal evidence, but unfortunately, few industry datasets exist to support it. Future work includes finding or creating such datasets.

\mypara{Evaluation on higher-end GPUs} In \S\ref{sec:eval}, we use NVIDIA A40 GPUs to conduct the experiments. 
We acknowledge that with very high-power GPUs and relatively low bandwidth, \name might not significantly improve over the text context baseline. 
Furthermore, due to GPU memory limitations, we have not evaluated our ideas on extra-large models such as OPT-175B. 
Evaluating \name on more powerful GPUs and larger LLMs is left for future work.

\mypara{Other system designs} \S\ref{sec:design} covers \name's encoder and streamer design. Other aspects such as which storage device(s) to store KV cache, caching policies, and locating KV cache quickly are discussed in concurrent works~\cite{yao2024cacheblend, gao2024attentionstore, jin2024ragcache}. 
We leave combining \name with these works to future work.

\mypara{Other limitations}
Task-wise, we did not extensively evaluate \name's performance on ``free-text generation'' tasks such as story generation because the quality metrics are less well-defined than the tasks in our evaluation.   
Network-wise, our network model does not include conditions with extremely high bandwidths.
Additionally, not all LLM applications can cache KV features.
Search-based apps, like Google and Bing, use real-time search results as context, and their volatile contexts will unlikely be reused unless for very popular search results. 
We expect future work to address these issues.

}




\tightsection{Conclusion}
We present \name, a context-loading module to minimize overall delays in fetching and processing contexts for LLMs.
\name reduces the bandwidth needed to transmit long contexts' KV cache through an encoder tailored to compress KV cache into compact bitstreams.
Experiments across three models of various capacities and four datasets with various context lengths show that \name reduces overall delays while maintaining high task performance.

\section*{Acknowledgement} 

We thank all the anonymous reviewers and our shepherd, Chen Qian, for their insightful feedback and suggestions.
The project is funded by NSF CNS-2146496, CNS-2131826,
CNS-2313190, CNS-1901466, CNS-1956180, CCF-2119184,  UChicago CERES Center, and Marian and Stuart Rice Research Award. The project is also supported by Chameleon Projects~\cite{chameleon}.
\pagebreak

\bibliographystyle{ACM-Reference-Format}
\bibliography{citations} 



\pagebreak
\appendix

\label{sec:appendix}

Note: Appendices are supporting material that has not been peer-reviewed.
\section{Text Output Examples of CacheGen}
\label{app:output}

\begin{figure}[h]
    \centering
    \includegraphics[width=0.99\columnwidth]{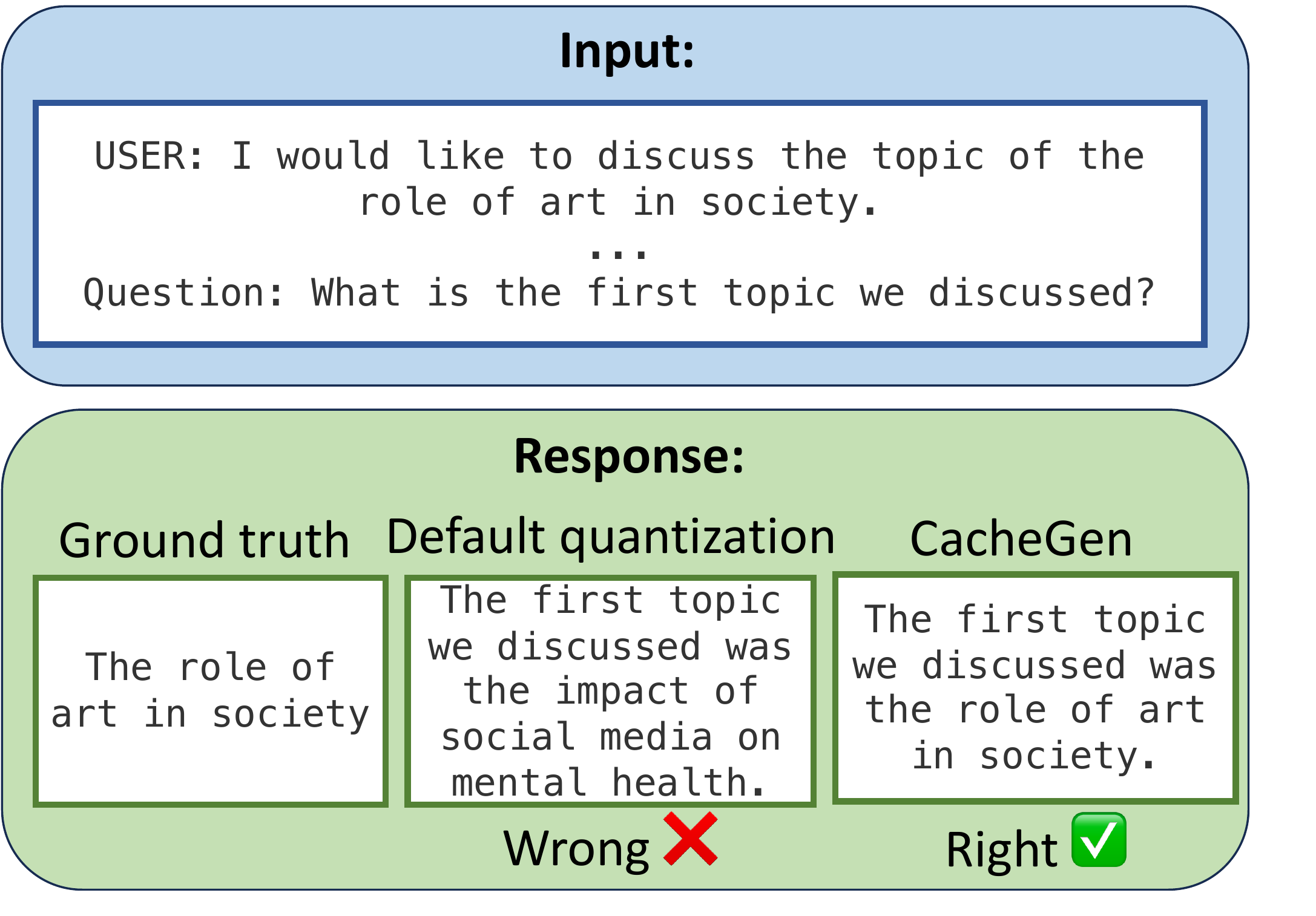}
    \tightcaption{An example of \name's output on the LongChat dataset with LongChat-7b-16k model. }
    \label{fig:example}
\end{figure}

Figure~\ref{fig:example} visualizes an example from the LongChat dataset~\cite{longchat2023} used in \S\ref{subsec:overall_eval}. 
The context fed into the LLM is a long, multi-round conversation history between the LLM and the user. 
An abridged context is shown in the upper box, where the first topic is about the role of art in society. 
The prompt to the LLM asks ``What is the first topic we discussed? ''
\name correctly generates the answer, whereas the default quantization baseline, which has a similar compressed KV cache size as \name, generates the wrong answer.

\begin{figure*}[t]
\centering
    \includegraphics[width=1.9\columnwidth]{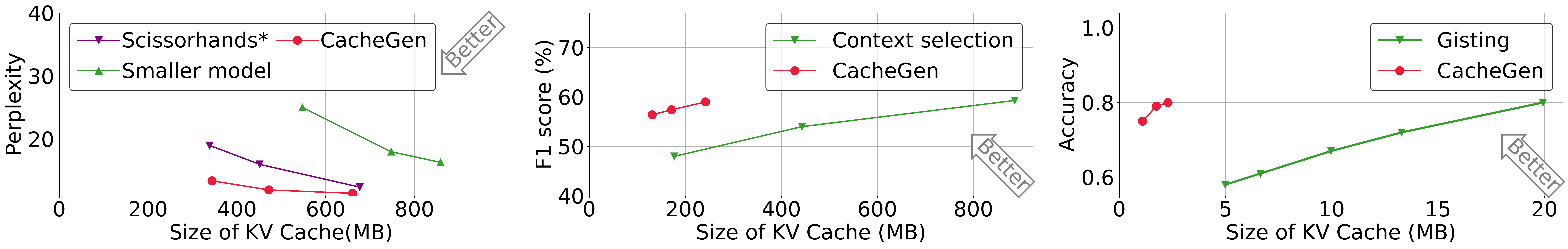}
    \tightcaption{Comparing \name and more intrusive methods, including smaller models, token dropping (left), context selection (middle), and gisting (right).
    }
    \label{fig:intrusive}
\end{figure*}


\section{\name vs. more intrusive methods}
\label{appendix:intrusive}
So far, all methods we have evaluated, including \name, do not modify the LLM or context.
As a complement, Figure~\ref{fig:intrusive} tests \name against recent methods that {\em change} the context or LLM. 
\begin{packeditemize}
\item {\em Smaller models:} 
Replacing the LLM with {\em smaller models} may speed up the computation.
Figure~\ref{fig:intrusive}a replaces the Llama-7B model with a smaller Llama-3B and applies different quantization levels. 
\item {\em Token selection:}
Figure~\ref{fig:intrusive}b uses \scis as an example of {\em removing tokens} with low self-attention scores from the LLM input~\cite{liu2023scissorhands}. 
Since the self-attention scores are only available during the actual generation, it cannot reduce TTFT, but we make an effort to create an idealized version of \scis (Scissorhands*) by running the self-attention offline to determine which tokens to drop and provide this information to Scissorhands* online.
\item {\em Gisting}
Finally, we test Gisting as an example of a more advanced method that shortens contexts into gist tokens and changes the LLM to accept the gist tokens~\cite{gisting}.
\edit{In Figure~\ref{fig:intrusive}c, we test the pre-trained gisting model, which is based on Llama-7B. 
The gisting model retrains the LLM's attention model in order to run inference on a compressed version of the input prompts. 
Since the gisting model can compress arbitrary long contexts into \textit{one token}, we vary the compression ratio of the gisting model to obtain a trade-off in size and accuracy.
This is done by adapting the fraction of input tokens that are compressed into one token.
We apply \name on the original Llama-7B model on the PIQA~\cite{bisk2019piqa} dataset, which is one of the most popular question-answering datasets. 
We did not apply \name on other datasets in our evaluation because the public pre-trained gisting model can only take up to 512 tokens, and truncating the dataset into smaller will not be able to preserve the information in the context. 
}
\end{packeditemize}
We can see that \name outperforms these baselines, reducing TTFT or KV cache size while achieving similar or better LLM's performance on the respective tasks.
In particular, \name is faster than smaller models (which are slowed down by transformer operations), and can reduce KV cache better than context selection or gisting because it compresses the KV features to more compact bitstream representations.
We want to stress that even though \name is compared head-to-head with these methods, it makes no assumption about the context and the model, so one can combine \name with these methods to potentially further improve the performance.
\clearpage
\section{\name System Settings}
\label{app:algorithm}
\subsection{KV Streamer Adaptation Logic}
We present the pseudo-code for the KV streamer logic that adapts to bandwidth here.
\label{Adaptation Logic}

\begin{algorithm}
\caption{\name Streaming Adapter Logic}\label{alg:adapter}

\begin{algorithmic}

\STATE $chunks\_to\_send \gets$ context chunks

\WHILE{$chunks\_to\_send \neq empty$}
\STATE get $chunk\_data$
\STATE $throughput \gets$ network throughput 
\STATE $remaining\_time \gets SLO - time\_elapsed$  

\IF{$time\_recompute \leq remaining\_time$ }
\STATE $cur\_chunk \gets$  text of $chunk\_data$
\ELSE
\STATE $level \gets max(level|size(chunks\_to\_send, level) \div throughput \leq remaining\_time$
\STATE $cur\_chunk \gets encode(chunk\_data, level)$
\ENDIF
\STATE send $cur\_chunk$
\STATE $chunks\_to\_send \gets chunks\_to\_send \setminus chunk\_data$

\ENDWHILE
\end{algorithmic}
\end{algorithm}
\subsection{Default Encoding Level}
\label{medium_encoding}
By default, \name encoding is done with the following parameters: we partition the layers in the LLM into three groups with equal distance, and set quantization bins to be 0.5, 1, 1.5 respectively.

\section{\name's improvement under various workloads}
\label{app:space}

\begin{figure}[t]
    \centering
    \includegraphics[width=0.99\columnwidth]{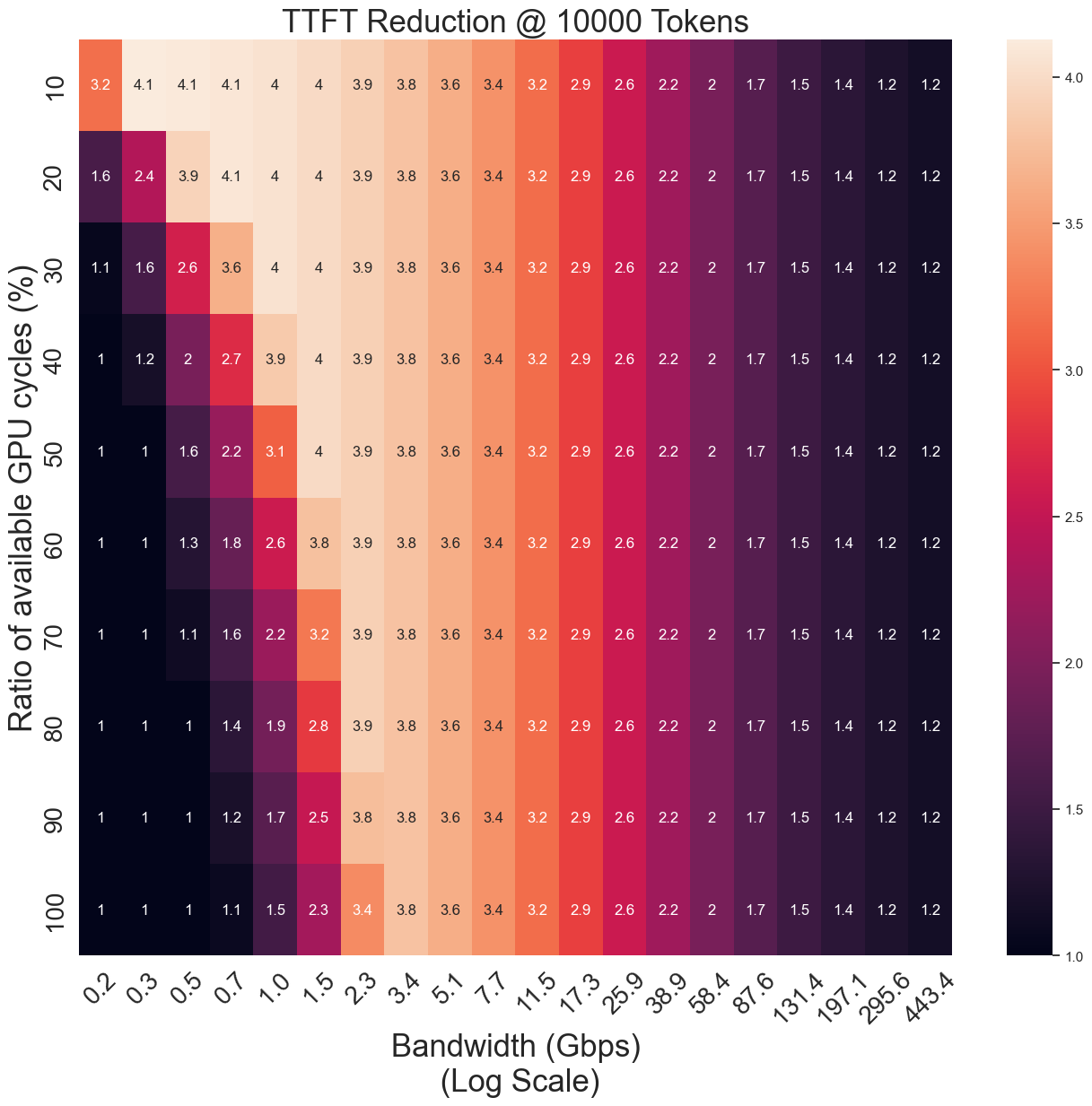}
    \tightcaption{Heatmap showing \name's improvement over the best baseline over a complete space of workloads. Brighter cells means more TTFT reduction.}
    \label{fig:heatmap-10k}
\end{figure}

Figure~\ref{fig:heatmap-10k} shows \name's improvement over the best baseline (between quantization and text context) over a complete space of workloads characterized along the two dimensions of GPU available cycles (\,  i.e., 1/$n$ with $n$ being the number of concurrent requests) and available bandwidth (in log scale). 
Figure~\ref{fig:sensitivity-bw} and Figure~\ref{fig:sensitivity-gpu} can be seen as horizontal/vertical cross-sections of this figure. 

\section{Cost of storing KV cache}
\label{sec:cost}
Our main focus in this paper is to reduce TTFT to achieve service SLO with minimal impact on the generation quality of LLM. However, context loading systems, especially \name, could be an economical choice for LLM service providers as well. For example, one piece of 8.5K-token context in Llama-13B takes roughly 5GB to store different versions compressed with \name. It costs \$0.05 per month to store this data on AWS~\cite{storage_pricing}. On the other hand, recomputing the KV cache from text costs at least \$0.00085 (input only) every time~\cite{together_pricing, anyscale_pricing, amazon_pricing, replicate_pricing}. If there are more than 150 requests reusing this piece of context every month, \name will also reduce the inference cost. The calculation here only serves as a rough estimation to highlight \name's potential. We leave the design of such a context loading system targeting cost-saving to future work.

\end{document}